\documentclass[pra, aps,
twocolumn,
amsmath,amssymb
,floatfix
]{revtex4-1}

\usepackage{physics}
\usepackage{epsf}
\usepackage{float}
\usepackage{graphicx}
\usepackage{amsmath}
\usepackage[usenames]{color}
\usepackage{float}
\usepackage{bm}
\usepackage{relsize}
\usepackage[normalem]{ulem}

\def \be {\begin{equation}}
\def \ee {\end{equation}}
\def \bea {\begin{align}}
\def \eea {\end{align}}
\def \p {\partial}
\def \BEA {\begin{eqnarray}}
\def \EEA {\end{eqnarray}}
\def \BC {\begin{cases}}
	\def \EC {\end{cases}}

\usepackage{siunitx}
\usepackage{tabularx}
\usepackage{xcolor}
\usepackage[version=4]{mhchem}
\usepackage{ctable}

\usepackage[unicode=true,colorlinks=true,citecolor=blue,urlcolor=blue]{hyperref}

\def \be {\begin{equation}}
\def \ee {\end{equation}}
\def \bea {\begin{align}}
\def \eea {\end{align}}
\def \p {\partial}
\def\bee{\begin{eqnarray}}
\def\eee{\end{eqnarray}}
\def \BC {\begin{cases}}
	\def \EC {\end{cases}}
\def \m {\bm }

\newcommand{\sdg}[1]{\textcolor{black}{#1}}

\newcommand{\Degree}[1]{\SI{#1}{\degree}}

\newcommand{\NanoMeter}[1]{\SI{#1}{\nano\meter}}

\newcommand{\MilliWatt}[1]{\SI{#1}{\milli\watt}}

\newcommand{\Hertz}[1]{\SI{#1}{\hertz}}
\newcommand{\MicroAmpere}[1]{\SI{#1}{\micro\ampere}}

\newcommand{\Kelvin}[1]{\SI{#1}{\kelvin}}

\begin{document}

\title{
 Ratchet effect in spatially modulated bilayer graphene:
 Signature  of  hydrodynamic   transport
}

\author{E. M{\"o}nch$^1$, S.~O. Potashin$^2$,  K.~Lindner$^1$, I. Yahniuk$^3$, L.~E. Golub$^2$, V.~Yu. Kachorovskii$^{2,3}$, V.~V~Bel'kov$^{2}$, 	
R.~Huber$^1$, K. Watanabe$^4$, T. Taniguchi$^5$, J. Eroms$^1$, D. Weiss$^1$,  and S.~D.~Ganichev$^{1,3}$}

\affiliation{$^1$Terahertz Center, University of Regensburg, 93040 Regensburg, Germany}

\affiliation{$^2$Ioffe Institute, 194021 St. Petersburg, Russia}

\affiliation{$^3$CENTERA, Institute of High Pressure Physics PAS, 01142 Warsaw, Poland}

\affiliation{$^4$Research Center for Functional Materials, 	National Institute for Materials Science, 1-1 Namiki, Tsukuba 305-0044, Japan,}

\affiliation{$^5$International Center for Materials Nanoarchitectonics, National Institute for Materials Science,  1-1 Namiki, Tsukuba 305-0044, Japan}

%

\date{\today, v.3}

\begin{abstract}
	 We report on the observation of the ratchet effect --- generation of  direct electric current in response to external terahertz (THz) radiation --- in bilayer graphene,
	 	  where   inversion symmetry is  broken by an  asymmetric   dual-grating  gate potential.  
	 	 	 As a central result, we demonstrate that at high temperature, $T = 150$~K, the ratchet current decreases at high frequencies as $ \propto 1/\omega^2$, while at low temperature, $T = 4.2$~K, the frequency dependence becomes much stronger $\propto 1/\omega^6$. The developed  theory shows that the frequency dependence of the ratchet current is very sensitive to the ratio  of the electron-impurity and electron-electron scattering rates. The theory predicts  that the dependence $1/\omega^6$ is realized in the hydrodynamic regime, when electron-electron scattering dominates, while $1/\omega^2$ is specific for the drift-diffusion approximation. Therefore, our experimental observation of a very strong frequency dependence reveals the emergence of the hydrodynamic regime.

\end{abstract}

\maketitle

\section{Introduction}
\label{intro}

Electronic fluid dynamics is one of the extremely actively developing areas of condensed matter physics (for review, see, e.g., Refs.~\cite{Narozhny2017, Lucas2018,Narozhny2019, Polini2020}). Although the pioneering works~\cite{Gurzhi1963, Gurzhi1965, Gurzhi1968, Jong1995} on hydrodynamic electron and phonon transport
have been done a very long time ago, the topic did not generate much interest until recently.
The interest on hydrodynamic transport was triggered by the fabrication of ultraclean ballistic structures, primarily based on one-dimensional and two-dimensional carbon materials. Convincing manifestations of hydrodynamic behavior in the different transport regimes have been demonstrated in a number of recent experiments~\cite{Bandurin2016, Crossno2016, Moll2016, Ghahari2016, Kumar2017, Bandurin2018b, Braem2018, Jaoui2018, Gooth2018, Berdyugin2019, Gallagher2019, Sulpizio2019, Ella2019, Ku2020, Raichev2020, Gusev2020, Geurs2020, Kim2020, Vool2020, Gupta2021, Gusev2021, Zhang2021, Jaoui2021}. Moreover, literally in recent years, it has been possible to experimentally
visualize the hydrodynamic flow  in ballistic 2D systems by  using various nanoimaging techniques~\cite{Braem2018, Sulpizio2019,Ella2019, Ku2020, Vool2020}.

The purpose of the current work is to demonstrate the transition from the drift-diffusion (DD) regime, where scattering by disorder dominates, to the hydrodynamic regime (HD), where electron-electron scattering prevails,
in a single system, on one and the same experimental sample, 
with a smooth change in some external control parameter.
While direct current transport measurements have been the focus so far, we show here that photovoltaic measurements offer additional opportunities for exploring this transition.
In particular, as we will show in this work, the obtained frequency dependence of the photoresponse allows one to extract important information about the type of transport in the system. We will demonstrate that the radiation frequency dependencies of the DD and HD responses are strikingly different. We also find that the transition between these two regimes
can be realized by
changing the carrier density of the electronic system via gate voltage.

In order to demonstrate DD-HD transition experimentally in photovoltaics, we choose one of the most general and fascinating phenomena in optoelectronics, which is the ratchet effect --- the generation of a dc electric current in response to an ac electric field in systems with broken inversion symmetry, for reviews see, e.g., Refs.~\cite{Linke2002,Reimann2002,Haenggi2009,Ivchenko2011,Denisov2014,Bercioux2015,Cubero2016,Budkin2016,Reichhardt2017,Ganichev2017}.
This general definition can be used for long periodic grating gate structures
with an asymmetric configuration of gate electrodes, e.g.,
dual-grating top gate (DGG) structures~\cite{Olbrich2011,Otsuji2013,BoubangaTombet2014,Faltermeier2015,Olbrich2016}.  In this case, the direction of the current is controlled by the lateral asymmetry parameter~\cite{Budkin2016,Faltermeier2017}
%
\begin{equation}
\label{Xi}
\Xi = \overline{|\bm E(x,t)|^2 {\mathrm{d}U(x)\over \mathrm{d}x}},
\end{equation}
where $x$ is the coordinate in the direction perpendicular to the grating, the overline stands for the average over the ratchet period and time, $\mathrm{d}U(x)/\mathrm{d}x$ is the derivative of the electrostatic potential of the grating $U(x)$ with respect to the coordinate $x$ and $\bm E(x,t)$ is the radiation electric field being coordinate dependent due to the near-field diffraction.

The ratchet effect  was treated theoretically and observed experimentally in various low dimensional structures~\cite{Olbrich2011,Otsuji2013,BoubangaTombet2014,Faltermeier2015,Olbrich2016,Faltermeier2017,Olbrich2009a,Popov2011,Kannan2011,Nalitov2012,Kannan2012,Kurita2014,Rozhansky2015,Bellucci2016,Fateev2017,Faltermeier2018,Rupper2018,Yu2018,Hubmann2020,DelgadoNotario2020,Sai2021}, so that the ratchet current measurements can already be considered a standard tool.
Despite a large number of publications on the ratchet effect, the role of electron-electron collisions, which can drive the system into the hydrodynamic regime, has not been studied thoroughly.   
This is the central question on which we focus in this work.

\begin{figure}
	\centering
	\includegraphics[width=\linewidth]{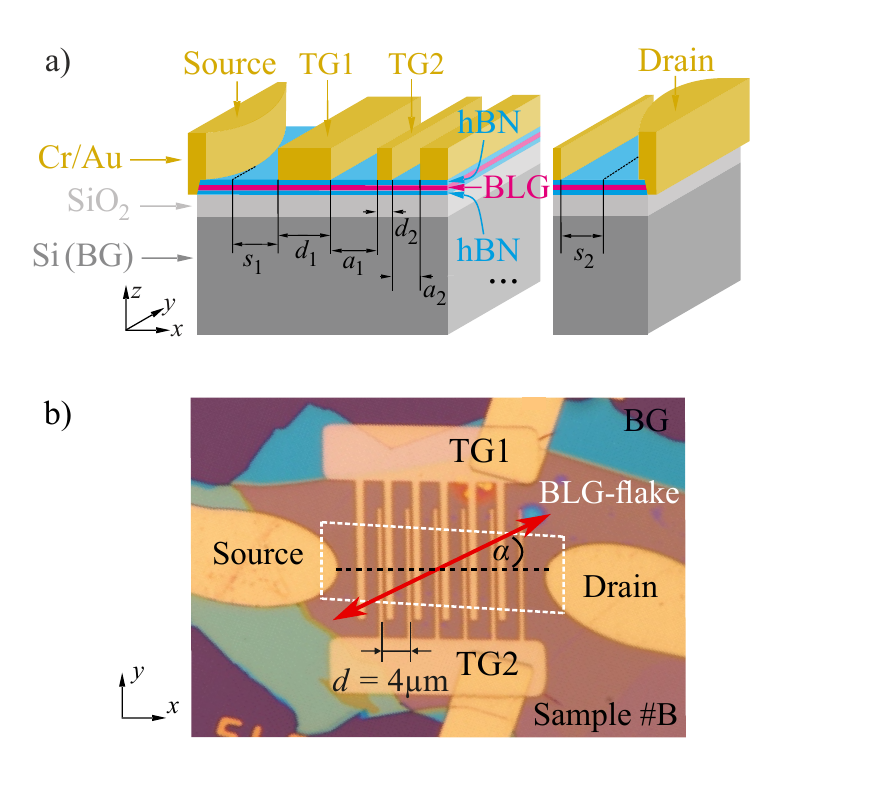}
	\caption{
		Cross-section (a) and optical micrograph (b) of bilayer graphene encapsulated by hBN layers with an inter-digitated lateral superlattice on top. Panel (a): The cross-section shows the layer sequence and indicates the width of the gold stripes ($d_{\rm 1,2}$) and the spacing in-between ($a_{\rm 1,2}$). The length of unstructured parts of the BLG flake between the source (drain) and the first (last) superlattice stripe is labeled as $s_{\rm 1}$ ($s_{\rm 2}$).The asymmetric supercell is repeated 6~times to create a superlattice with a period $L = d_{\rm 1} + d_{\rm 2} + a_{\rm 1} + a_{\rm 2}$. Panel (b): The interconnected thick and thin stripes form top gates TG1 and TG2, respectively. The white dashed tetragon shows the border of the BLG flake.
		Red arrow illustrates the radiation electric field vector $\bm E$ for normal incident linearly polarized radiation tilted from the $x$-axis by the azimuth angle $\alpha$.}
	\label{FigS1}
\end{figure}

Here, we demonstrate, both theoretically and experimentally, that the dc response of a DGG, based on bilayer graphene, has very different frequency dependencies: $1/\omega^6$ in the hydrodynamic regime and $1/\omega^2$ within the drift-diffusion approximation. We derive an analytical formula which describes the transition between both regimes and demonstrate that at
low carrier densities the experimental results follow the HD approach.
However, after increasing the sample's electron concentration via the back gate voltage the obtained data are in accordance with the description in the drift-diffusion picture.  
This suggests that we can tune between both regimes.
We further find that a temperature increase at sufficiently low temperatures, where the contribution of phonons is negligible,
shifts the system closer to hydrodynamic regime.

\section{Samples and methods}
\label{samples-methods}

\subsection{Samples}

Bilayer graphene (BLG) samples were encapsulated between hexagonal boron nitride (hBN) flakes to ensure a high sample quality and protect the active layer from influences of the environment. The heterostructures were fabricated by a van der Waals stacking technique \cite{Wang2013} on top of a Si wafer, serving as a uniform back gate, covered with $\NanoMeter{285}$ thermal SiO$_2$. The inter-digitated
DGG
structures were fabricated on top of encapsulated bilayer graphene by electron beam lithography (EBL), evaporation of 5 nm Cr and 30 nm Au, and lift-off.  Afterwards, source and drain contacts were prepared by EBL, reactive ion etching to expose the graphene layer, and evaporation of Cr and Au. These contacts allow us to study the ratchet current generated in the direction perpendicular to the DGG stripes

A cross-section sketch and an optical micrograph
are shown in Figs.~\ref{FigS1}(a) and (b), respectively. The lateral superlattice consists of two top gates which comprise six cells. 
Both top gates, TG1 (wide stripes) and TG2 (narrow stripes), have different width and spacing parameters with a characteristic cell period $L$, listed in Tab.~\ref{TabS1}. In addition, they are electrically isolated from each other, which allows the application of unequal top gate voltages $U_\mathrm{TG1}$ and $U_\mathrm{TG2}$
providing a tunable asymmetric electrostatic potential, and, consequently controllable variation of the lateral asymmetry parameter $\Xi$. 

\begin{table*}
	\centering
	\begin{tabularx}{\textwidth}{XXXXX}
		\toprule[0.05cm]\addlinespace[0.2cm]
		Parameter&& Sample \#A 	& 	Sample \#B 		& 	 Sample \#C \\\midrule[0.025cm]\addlinespace[0.1cm]
		flake length / width &($\SI{}{\micro\meter}$) 	& 25 / 13 	&  30 / 11.5	& 17 / 6.5	\\
		top / bottom thickness of hBN& ($\SI{}{\nano\meter}$) & 30 / 60 	&  40 / 80	& 55 / 70	\\
		\midrule[0.015cm]\addlinespace[0.1cm]
		$a_1$ / $a_2$ &($\SI{}{\micro\meter}$)	& 2 / 0.5 	&  	2 / 0.5	& 1 / 0.25	\\
		$d_1$ / $d_2$& ($\SI{}{\micro\meter}$)	& 1 / 0.5 	&  1 / 0.5 	& 0.5 / 0.25	\\
		$s_1$ / $s_2$& ($\SI{}{\micro\meter}$)	& 1.4 / 0.6 	& 3 / 3.4 	& 2.3 / 3.1	\\
		\addlinespace[0.1cm]\bottomrule[0.05cm]
	\end{tabularx}	
	\caption{Geometric parameters of the samples \#A, \#B, and \#C. For the structure cross-section and top view see Fig.~\ref{FigS1}.
	}
	\label{TabS1}
\end{table*}

\subsection{Methods}

The ratchet current in the bilayer graphene samples was driven by in-plane alternating electric fields $\bm{E}(t)$ of radiation provided by a continuous wave ($cw$) optically pumped molecular gas laser~\cite{Kvon2008,Dantscher2017}. In the experiments described below we used radiation with frequencies
$f= 2.54$, 1.63, and 0.69~THz (corresponding photon energies $\hbar\omega = 10.5$, 6.7, and 2.9~meV, respectively).
The incident power, $P$, lying in the range from 15 to $\MilliWatt{80}$, was modulated at a frequency of 60~Hz by a mechanical chopper. To control the
laser power stability during the measurements it was monitored by reference pyroelectric detectors.
The beam cross-section was controlled by a pyroelectric camera revealing a nearly Gaussian profile with a spot diameter which, depending on the frequency, ranges from 1.5 to 3~mm at sample's position.
Consequently, the radiation intensity was reaching $I \approx \SI{3}{\watt\centi\meter^{-2}}$ and the THz electric field $E \approx 50$~V/cm.


To extract and study different ratchet effects, such as the Seebeck, linear, and circular ratchets,
we make use of their different polarization dependencies, see Refs.~\cite{Ivchenko2011,Olbrich2016} and discussion below. A controllable variation of the radiation's polarization state was obtained rotating lambda-half or lambda-quarter crystal quartz wave plates. In the former case, the orientation of the linearly polarized radiation is described by the azimuth angle $\alpha$ between the electric field vector $\bm E$ of the radiation and the $x$-axis.
The change of the radiation helicity 
is defined by the angle $\varphi$ between the initial polarization plane $E_l$ and the optical axis of the lambda-quarter plate. Note that for $\alpha = 0$ and $\varphi = 0$ the electric field is directed along the $x$-axis, i.e., perpendicular to the stripes, see inset in Fig.~\ref{FigS1}(b). 
The samples were illuminated through $z$-cut crystal quartz windows of a temperature-regulated Oxford Cryomag optical cryostat. The windows were covered by black polyethylene films, which are transparent for terahertz radiation, but prevent uncontrolled illumination of the sample with room light in both the visible and the infrared ranges.

\begin{figure*}
	\centering
	\includegraphics[width=0.9\linewidth]{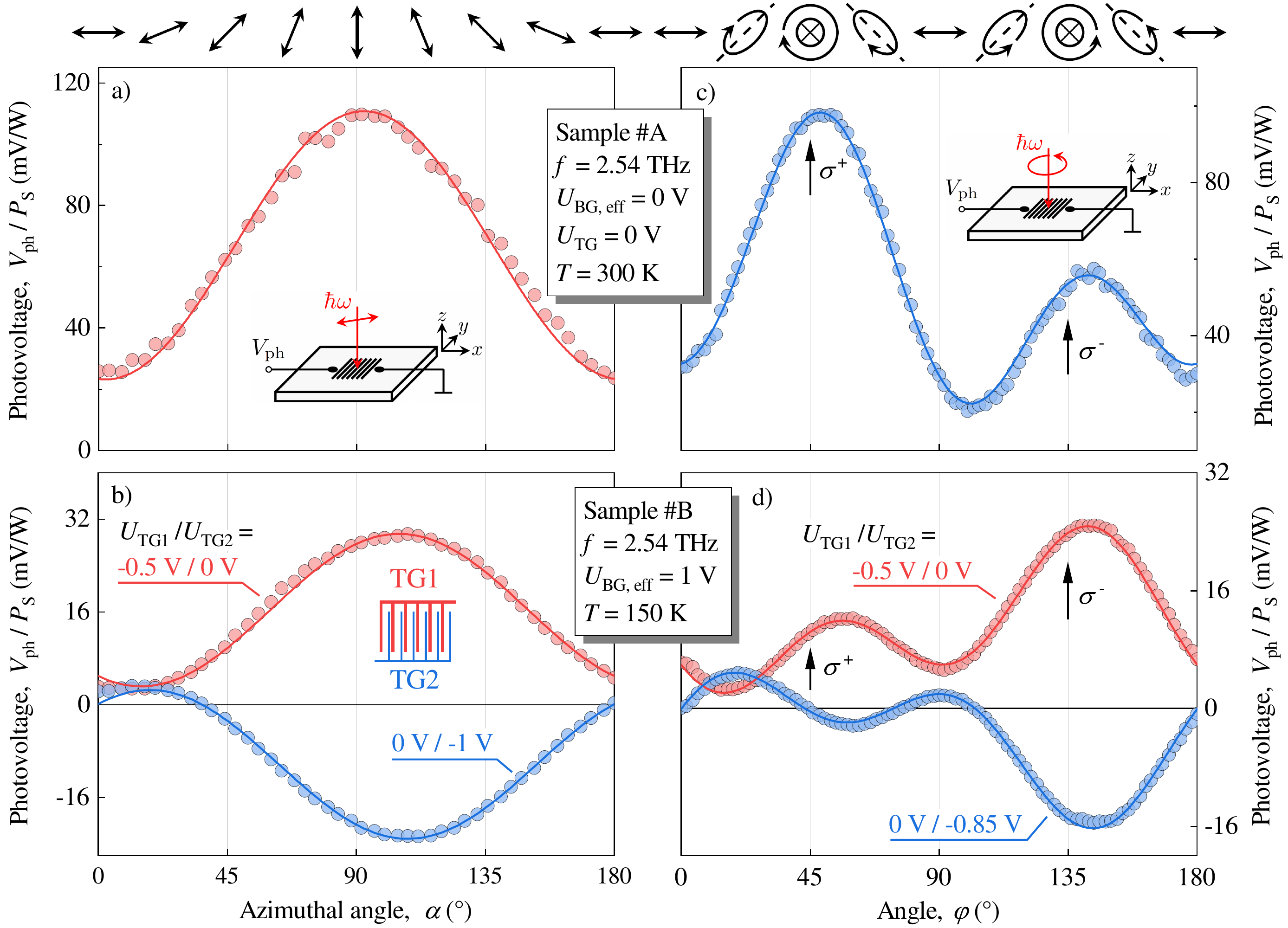}
	\caption{Polarization dependencies of the photovoltage $V_\mathrm{ph}$ normalized to radiation power $P_\mathrm{S}$ measured at $f = 2.54$~THz. Data presented in panels (a) and (b) are obtained by rotation of a lambda-half plate, demonstrating  variation of the signal upon rotation of the radiation electric field vector $\bm E$ in respect to $x$-axis. For azimuth angles $\alpha = 0$ and \Degree{180} $\bm E$ is normal to the top gate stripes, whereas for $\alpha = \Degree{90}$ the field is oriented along the stripes. Arrows on top illustrate orientations of radiation electric field vector for several values of $\alpha$. Data presented in panels (c) and (d) are obtained by rotation of a lambda-quarter plate by an angle $\varphi$ which allows us to examine  radiation helicity dependence of the photovoltage. Upward arrows mark right-($\sigma^+$) and left-($\sigma^-$) handed circularly polarized radiation. The ellipses on top illustrate the polarization states at several angles $\varphi$. Solid curves are fits after Eq.~(\ref{linear}), panels (a) and (b), and Eq.~(\ref{circular}), panels (c) and (d), see also theoretical Eqs.~\eqref{j-hyd-x}-\eqref{j-dr-y}. The data in panels (a) and (c) are obtained for sample \#A at room temperature connecting both top gates to ground. The data in panels (b) and (d) present photovoltage generated in sample \#B at $T=150$~K for $U_{\rm BG, eff} =1$~V and for lateral asymmetries of the applied electrostatic potential with opposite signs.
		The insets in panels (a) and (c) represent the measurement configurations of the sample for linearly and elliptically polarized radiation. The inset in panel (b) shows the DGG stucture. Here, the color code corresponds to the applied top gates.}
	\label{FigR1}
\end{figure*}

The photovoltage signal was measured as a voltage drop, $V_\text{ph}$, directly over the sample resistance, $R_\mathrm{s}$, applying a standard lock-in technique.
In all graphs, the photovoltage is normalized to the radiation power coming onto the sample, $P_{\rm S}$, with  $P_\mathrm{S} = IA_\mathrm{S}$, where $I$ is the radiation intensity and $A_\mathrm{S}$ defines the area of the DGG on top of the bilayer flake. Note that the corresponding photocurrent $J_{\rm dc}$ relates to the photovoltage $V_\text{ph}$ as $J_{\rm dc} = V_\text{ph} / R_\mathrm{s}$.

\section{Results}
\label{results}

Illuminating the bilayer graphene superlattice we observed
a photosignal exhibiting the characteristic behavior of the ratchet effect. This includes the dependence on the lateral asymmetry parameter $V_\text{ph}\propto \Xi$ and the variation of the ratchet current with the polarization of the radiation~\cite{Ivchenko2011,Olbrich2016}.
%
Figures~\ref{FigR1}(a) and (b) exemplarily show the dependence of the photovoltage
on the orientation of the electric field vector $\bm E$ of the linearly polarized radiation passing through a lambda-half plate. 
The data can be well fitted by
\begin{equation}
V_\mathrm{ph}(\alpha) = V_0 + V_\mathrm{L1}\cos2\alpha + V_\mathrm{L2}\sin2\alpha,
\label{linear}
\end{equation}
where the terms  $\cos2\alpha = P_{\rm L1}$ and $\sin2\alpha = P_{\rm L2}$ represent the Stokes parameters
and correspond to the linear polarization degrees for the axes $x, y$ and in the coordinate system rotated by an angle of \Degree{45}, respectively~\cite{Saleh1991,Belkov2005}. Note that the amplitude $V_\mathrm{L1}$ was more than an order of magnitude greater than $V_\mathrm{L2}$ for all studied conditions. The observed polarization dependence is expected from the phenomenological theory of the ratchet effect presented in Refs.~\cite{Ivchenko2011,Olbrich2016} demonstrating that the fitting parameters $V_0, V_\mathrm{L1}$, and $V_\mathrm{L2}$ describe the polarization-independent ratchet contribution ($V_0$) and the linear ratchet effect ($V_\mathrm{L1}, V_\mathrm{L2}$). Moreover, in this regard, the theory reveals that the ratchet contributions should reverse sign upon inversion of the lateral asymmetry. This feature is clearly demonstrated in Fig.~\ref{FigR1}(b), which shows the data for opposite signs of the parameter $\Xi$, obtained by using different combinations of the effective top gate voltages: $U_{\rm TG1}=-0.5$~V/ $U_{\rm TG2}=0$ (red circles) and $U_{\rm TG1}=0$/ $U_{\rm TG2}=-1$~V (blue circles). Note that for zero top gate voltages, see Fig.~\ref{FigR1}(a), the asymmetry is created by the built-in potential due to the metal stripes of different widths (TG1 and TG2) deposited on top of the encapsulated bilayer graphene.

Using  elliptically (circularly) polarized radiation
allows us to explore a further ratchet contribution, the circular ratchet effect~\cite{Ivchenko2011,Olbrich2016}, which is characterized by opposite signal polarities in response to the radiation of opposite helicities. The corresponding dependencies on the rotation angle, $\varphi$, of the lambda-quarter plate are shown in Figs.~\ref{FigR1}(c) and~(d). The data can be well fitted by
\begin{eqnarray}
V_\mathrm{ph}(\varphi) &=&  V_0 + V_\mathrm{L1} \left(\frac{\cos4\varphi + 1}{2}\right) + V_\mathrm{L2} \left(\frac{\sin4\varphi}{2}\right) + \nonumber \\ &+& V_\mathrm{C}\sin2\varphi.
\label{circular}
\end{eqnarray}
Here the terms in brackets describe the variation of the Stokes parameters
$P_{\rm L1}$ and $P_{\rm L2}$ in $\lambda$-quarter arrangement~\cite{Saleh1991,Belkov2005}, and the circular effect is given by the last term on the right hand side of Eq.~(\ref{circular}) with fitting parameter $V_\mathrm{C}$ and Stokes parameter $P_{\rm C} = \sin 2\varphi$, which determines the degree of circular polarization.
%
Also for the circular contribution to the total photosignal we observed the expected ratchet behavior $V_{\rm C} \propto \Xi$. This is exemplarily shown in Fig.~\ref{FigR1}(d), demonstrating that the polarity of the circular contribution $V_\mathrm{C}$ reverses by changing the sign of the lateral asymmetry parameter $\Xi$, obtained by different combinations of the top gate voltage.
Note that for $\varphi = \Degree{45}$ and $\Degree{135}$ the radiation is circularly polarized and, consequently, the linear ratchet contributions proportional to $V_\mathrm{L1}$ and $V_\mathrm{L2}$
vanish. The polarization behavior of the ratchet response addressed above, see Eqs.~(\ref{linear}) and (\ref{circular}), allows us to analyze all three ratchet contributions: the linear, polarization-independent, and the circular one.
By that the linear and polarization-independent contributions were extracted measuring the photoresponse $V_{\rm ph}(\alpha)$ to the linearly polarized radiation with azimuth angle $\alpha = 0$ and $90^\circ$, and using $V_{\rm 0} = [V_{\rm ph}(0) + V_{\rm ph}(90^\circ)]/2$ and $V_{\rm L1} = [V_{\rm ph}(0) - V_{\rm ph}(90^\circ)]/2$. Because of $V_{\rm L2} \ll V_{\rm L1}$, the discussion below of the linear ratchet effect is limited to $V_{\rm L1}$. The circular contribution $V_{\rm C}$ was obtained by measuring the ratchet current in response to circularly polarized radiation and using $V_{\rm C} = [V_{\rm ph}(45^\circ) - V_{\rm ph}(135^\circ)]/2$.

\begin{figure}
	\centering
	\includegraphics[width=\linewidth]{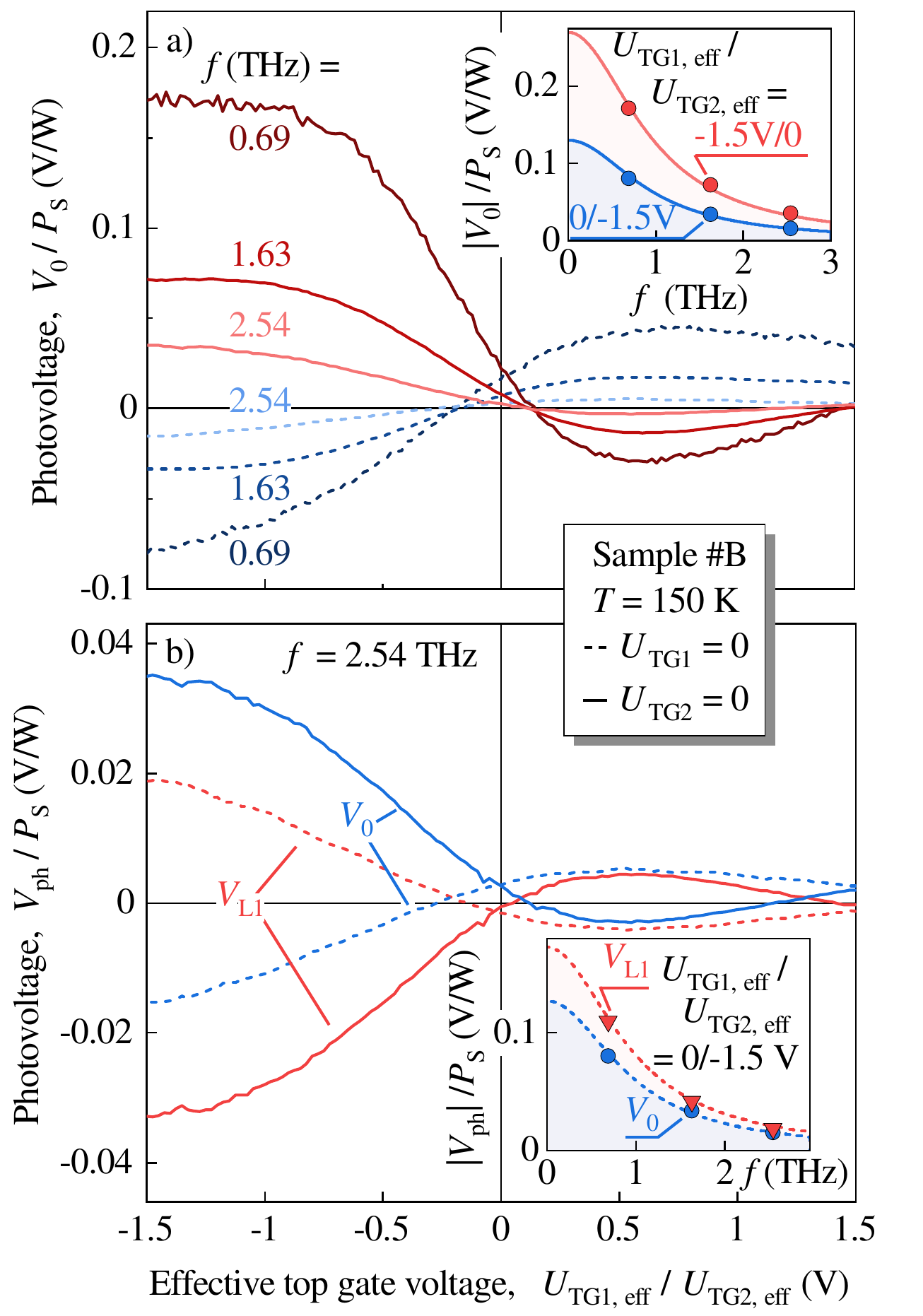}
	\caption{Top gate dependencies of the polarization-independent, $V_\mathrm{0}/P_\mathrm{S}$, and linear, $V_\mathrm{L1}/P_\mathrm{S}$, ratchet signals. The signals are normalized to the radiation power $P_{\rm S}$. The data were obtained sweeping one top gate potential and holding the other at ground. The individual contributions were extracted from the total photoresponse applying difference in their polarization dependencies, see Eq.~(\ref{linear}) and corresponding discussion. The data are presented for sample \#B, $T = 150$~K and $U_{\mathrm{BG, eff}} = 1$~V. Solid curves show dependencies on $U_\textrm{TG1}$ and the dashed curves dependencies on $U_\textrm{TG2}$. (a) $U_{\rm TG1}$-dependence of the polarization-independent ratchet signal $V_0/P_\mathrm{S}$ obtained for different radiation frequencies $f = 0.69$, 1.63, and 2.54 THz. The color code of the curves refers to the color of the labeled frequencies. (b) Comparison of the top gate dependencies of the linear ratchet $V_\mathrm{L1}/P_\mathrm{S}$ and polarization-independent ratchet $V_0/P_\mathrm{S}$ excited by radiation with $f = 2.54$~THz. The insets show the absolute value of the amplitudes as a function of the radiation frequency for different combinations of the top gate voltages. Solid and dashed lines in the insets are fits after Eq.~(\ref{Drude}) with 
	$\tau = 0.17$~ps.}
	\label{FigR4}
\end{figure}

To explore the dependence of the individual ratchet effects on the lateral asymmetry $\Xi$ we varied one top gate voltage ($U_{\rm TG1}$ or $U_{\rm TG2}$), while keeping the other at ground potential. Figure~\ref{FigR4} shows these dependencies obtained by applying radiation of different frequencies. First of all these figures clearly demonstrate that the inversion of $\Xi$, achieved either by changing the polarity of the gate voltage applied to one of the gates or by exchanging the biased gates, yields opposite signs of the photosignal. This behavior is shown for linear, $V_{\rm L1}$ and polarization-independent, $V_{\rm 0}$, contributions, see Figs.~\ref{FigR4}(a) and (b), respectively. The same results are obtained for all samples and all ratchet contributions, see Appendix~\ref{Additional_measurements}.

Analyzing the results for different frequencies we observe that the overall behavior of the ratchet current is the same and the only difference is a substantial increase of the current amplitude with decreasing frequency. This is shown in Fig.~\ref{FigR4}(a), presenting the top gate dependencies of the polarization-independent contribution, $V_0$, measured in sample \#B for three radiation frequencies at $T=150$~K.
Figure~\ref{FigR4}(b) shows that an increasing signal with the reduction of frequency is a distinctive feature for both,
polarization-independent and linear ratchet effect. The insets in Fig.~\ref{FigR4} show the frequency dependencies of the ratchet magnitude obtained for opposite signs of the lateral asymmetry $\Xi$ ($U_{\rm TG1, eff}/ U_{\rm TG2, eff} = -1.5$~V/ 0 and $0/-1.5$~V). Here the effective top gate voltages are indicated as $U_{\rm TG1/TG2, eff} = U_{\rm TG1/TG2} - U_\mathrm{TG1/TG2, CNP}$, where the $U_{\rm TG1/TG2, CNP}$ is the voltage value which corresponds to the charge neutrality point (CNP). For top gate dependencies of the resistivity see Appendix~\ref{Transport}.  Note that the signal changes its sign in the vicinity of zero effective top gate bias. Therefore, we analyzed the frequency dependency for high negative gate voltages at which the $U_{\rm TG}$-dependencies become almost flat. The obtained data can be well fitted by the Lorentz formula
\begin{equation}
V_\mathrm{ph}(f)/P_\mathrm{S} \propto \frac{V_{\rm ph}(0)}{1 + (2\pi f)^2\tau^2}
\label{Drude}
\end{equation}
with the fitting parameter $\tau$ and $V_{\rm ph}(0)$, corresponding to the zero frequency amplitude, see dashed and solid lines in the insets of Figs.~\ref{FigR4}(a) and (b). 
The best fits are obtained with the momentum relaxation time of 
$\tau = 0.17$~ps, which is several times shorter than that obtained from the width of the cyclotron resonance (CR) (not shown) studied in the same samples ($\tau = 0.6$~ps at 4.2~K). This difference is attributed to a substantially higher temperature at which the measurements in Fig.~\ref{FigR4} were performed ($T = 150$~K).

\begin{figure}
	\centering
	\includegraphics[width=\linewidth]{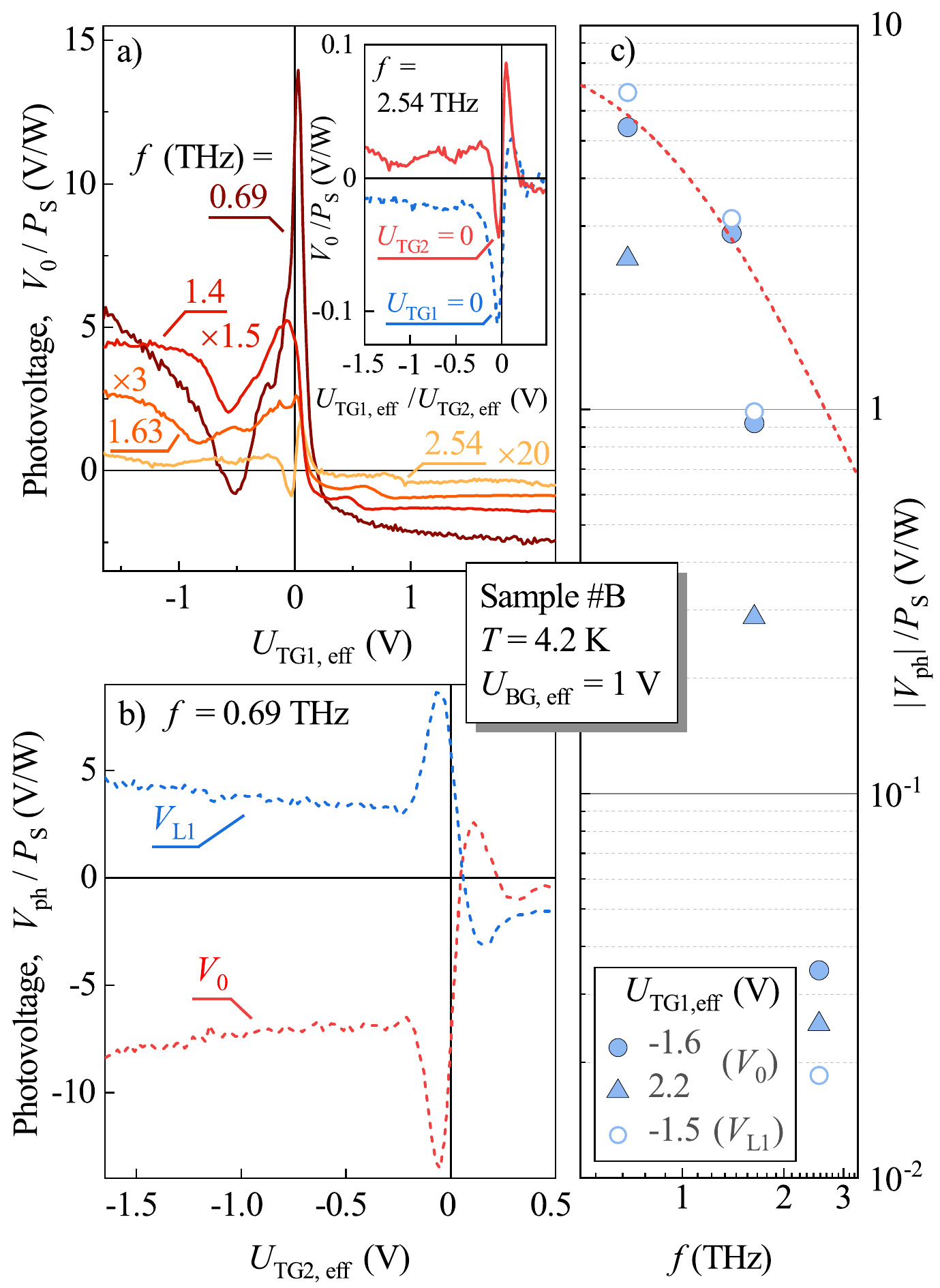}
	\caption{Top gate dependencies of the normalized ratchet signal obtained in sample \#B at low temperature, $T = 4.2$~K, applying radiation with different frequencies. The data are measured for $U_{\rm BG, eff} = 1$~V.
		Panel (a): polarization-independent contribution, $V_\mathrm{0}/P_\mathrm{S}$, as a function of the effective TG1 potential $U_{\rm TG1, eff} = U_{\rm TG1} - U_\mathrm{TG1, CNP}$. The inset shows TG1- and TG2-dependencies of the signal excited by radiation with $f = 2.54$~THz. Panel (b) shows polarization-independent, $V_\mathrm{0}/P_\mathrm{S}$, and linear, $V_\mathrm{L1}/P_\mathrm{S}$, ratchet contributions as a function of the effective TG2 potential $U_{\rm TG2, eff} = U_{\rm TG2} - U_\mathrm{TG2, CNP}$. Panel (c) shows the frequency dependence of both contributions for several values of $U_{\rm TG1. eff}$ using a double logarithmic scaling. The values of $U_{\rm TG1}$ used for this plot correspond to almost flat parts of the curves for all used frequencies. 
\sdg{The dashed red line is calculated after Eq.~(\ref{Drude}) with $\tau = 0.17$~ps, which was used to fit the data in the insets in Fig.~\ref{FigR4}}. It demonstrates that in contrast to the data for $T = 150$~K the frequency dependence of the ratchet signals at low temperatures is very strong.
	}
	\label{FigR5}
\end{figure}



Cooling the sample from 150 to 4.2~K on the one hand modifies the top gate dependencies in the vicinity of $U_{\rm TG1/TG2, eff} \approx 0$, and on the other hand drastically changes the signal magnitudes as well as their frequency dependencies, see Fig.~\ref{FigR5}. Apart of that, Figs.~\ref{FigR5}(a) and (b) clearly demonstrate the characteristic ratchet behavior $V_{\rm 0} \propto \Xi$: the photoresponse inverses its sign by exchanging $U_{\rm TG1}$ and $U_{\rm TG2}$ as well as changing the polarity of the electrostatic potential applied to one of the top gates.

At first we address the peak feature, which is clearly seen at $U_{\rm TG1/TG2, eff} \approx 0$. Figures~\ref{FigR5}(a) and~(b) show that under these conditions the top gate dependencies of the signals are in accordance with the first derivative of the conductance~\footnote{We note that such sign-alternating behavior in the vicinity of the CNP was also detected in the back gate voltage dependencies measured for zero top gate biases, see Appendix~\ref{Additional_measurements}. Similar effect was also previously reported for the ratchet effects in monolayer graphene~\cite{Olbrich2016}.} (for top gate dependence of the resistivity see Appendix~\ref{Transport}). This behavior is characteristic for both the polarization-independent and the linear ratchet effects, see Fig.~\ref{FigR5}(b).

\begin{figure*}
	\centering
	\includegraphics[width=0.9\linewidth]{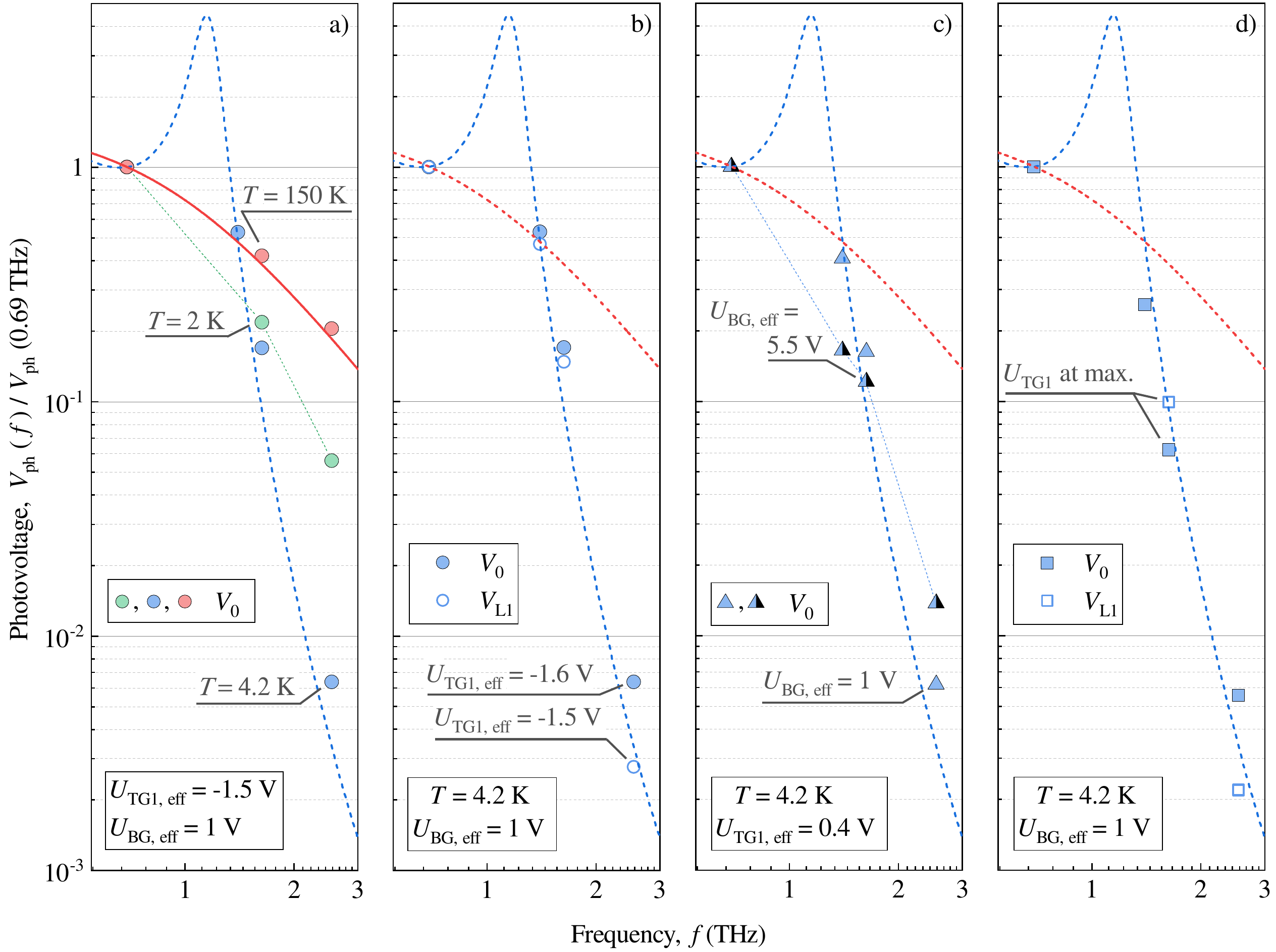}
	\caption{Frequency dependencies of the two ratchet contributions, $V_0$ and $V_{\rm L1}$, for different temperatures and effective back gate voltages measured for sample \#B. For comparison of the signal behavior upon the change of the radiation frequency each data set is normalized on the value of the signal at the lowest frequency ($f=0.69$~THz). For better visualization of the different frequency dependencies a double logarithmic scale was used to present the data. 
		\sdg{(a) Data for three temperatures, 2, 4.2 and 150~K, obtained for $U_{\rm BG, eff} = 1$~V and $U_{\rm TG1}=-1.5$~V. The red solid curve (displayed as red dashed curves in panel (b)-(d)) corresponds to Eq.~(\ref{Drude}) with $\tau = 0.17$~ps used to fit high temperature data in the insets in Fig.~\ref{FigR4}, whereas the dashed blue curve represents the crossover from DD to HD following Eq.~(\ref{final-current-large-s}).
			(b) Both contributions polarization-independent, $V_0$, and linear, $V_{\rm L1}$, for $U_{\rm BG, eff} = 1$~V and positive/negative values of the effective TG1 potential.  
			(c) $V_0$ for $T=4.2$~K and two back gate potentials. (d) $V_0$ and $V_{\rm L1}$ for $T=4.2$~K and $U_{\rm BG, eff} = 1$~V obtained at signal's extreme detected close to effectively zero top gate potentials. The thin linking lines (green and blue in panels (a) and (c), respectively) between the experimental points serve as guide lines for better illustration. The experimental data are shown for $U_{\rm TG2} = 0$}.}
	\label{FigR6}
\end{figure*}

Now we turn to the change of the signal magnitudes and, in particular, their frequency dependencies. Strikingly, the reduction of temperature from 150 to 4.2~K drastically changes the frequency dependence of the ratchet signal. While at a frequency of $f=2.54$~THz the temperature drop from 150 to 4.2~K does not affect much the photoresponse magnitude, at lower frequencies the decrease in temperature significantly (by two orders of magnitude) increases the ratchet current. Alike for the analysis of the data obtained at $T=150$~K, which were presented above, we focus below on the frequency dependencies of the signals obtained at high negative/positive top gate voltages, where the signals are almost voltage independent. Since the signal amplitude changes by nearly two orders of magnitude when varying the radiation frequency by about four times, the frequency dependence is presented on a double logarithmic scale. 
\sdg{The red dashed line} in Fig.~\ref{FigR5}(c) shows the frequency dependence of $V_{\rm ph}$
\sdg{calculated after} Eq.~(\ref{Drude}), \sdg{ which is} used to fit the data at $T=150$~K. 
It is seen that at low temperatures the data deviate significantly from the expected Lorentz-like behavior.

To explore the characteristic features of the spectacular modification of the ratchet effect we measured the top gate dependencies of \sdg{linear, $V_{\rm L1}$, and polarization-independent, $V_0$, ratchets} at different frequencies, temperatures, and back gate voltages. The data were used to obtain frequency dependencies of the photosignals, presented in Figs.~\ref{FigR5}(c) and \ref{FigR6}. 
In order to investigate the difference in the frequency behavior \sdg{obtained for different sets of parameters (top/bottom gate voltages and temperatures)} we normalized each data set to the magnitude of the signal  at the lowest frequency used in our experiments ($f=0.69$~THz). The results are demonstrated in Figs.~\ref{FigR6}(a)-(d). \sdg{As it is addressed above,} because of the huge difference in the signal magnitude we used a double logarithmic presentation. Furthermore, we included in all panels
\sdg{red curves (solid/dashed)} calculated after Eq.~(\ref{Drude}) with the momentum relaxation time, $\tau = 0.17$~ps, used to fit the data at $T = 150$~K in the insets in Fig.~\ref{FigR4}. 

\sdg{Figure~\ref{FigR6}(a)} shows the data at $4.2$~K and low back gate voltage ($U_{\rm BG, eff} = 1$V) previously presented in Fig.~\ref{FigR5}(c), but now, as 
\sdg{addressed above}, normalized to the response obtained at $f=0.69$~THz. 
\sdg{It can be clearly seen} \sdg{ that the frequency dependence of the ratchet magnitude obtained at 4.2~K (blue circles) \sdg{is different} from that detected for 150~K (red circles).} For $f = 1.63$~THz  
the \sdg{normalized} ratchet amplitude already \sdg{differs} \sdg{from the Lorentz-curve (red solid line)} by about 
\sdg{three times}. Further increase of the frequency results in an abrupt reduction of the magnitude. Now, at $f = 2.54$~THz, the signal deviates from the Lorentz-curve by more than 
\sdg{30} times. Such a 
\sdg{drastic change of the frequency dependence} of the ratchet current is observed for both $V_0$ and $V_{\rm L1}$, see Fig.~\ref{FigR6}\sdg{(b)}. 
\sdg{The decrease of temperature from 4.2 to 2~K, however, results in a substantially weaker frequency dependence, see Fig.~\ref{FigR6}(a). \sdg{Comparing the normalized signal's magnitude for $T = 2$~K (green circles) and for $T = 4.2$~K (blue circles) with the Lorentz-curve we obtained that at $f = 2.54$~THz the former one is reduced by three times, whereas the latter one by 30 times.} 
		%
	The weaker frequency dependence is also detected for $T=4.2$~K, but higher carrier densities obtained by increasing the back gate voltage, see Fig. \ref{FigR6}(c) showing the data for $U_{\rm BG, eff} = 1$~V (blue triangles) and 5.5~V (double colored triangles).} 
%
%
%
%
\sdg{And last but not least, in Figs.} \ref{FigR6}(a)-(c) 
\sdg{we show the} data taken at high positive/negative top gate voltages, where the signals are almost 
independent \sdg{of the top gate voltage.} 
\sdg{The strong} frequency 
\sdg{dependence at $T= 4.2$~K and low carrier density} has also been detected at signal's extreme obtained for top gates biased close to effectively zero potentials. This is shown in Fig.~\ref{FigR6}(d) for $V_0$ and $V_{\rm L1}$ measured for $U_{\rm BG}=1$~V.

\begin{figure}
	\centering
	\includegraphics[width=0.9\linewidth]{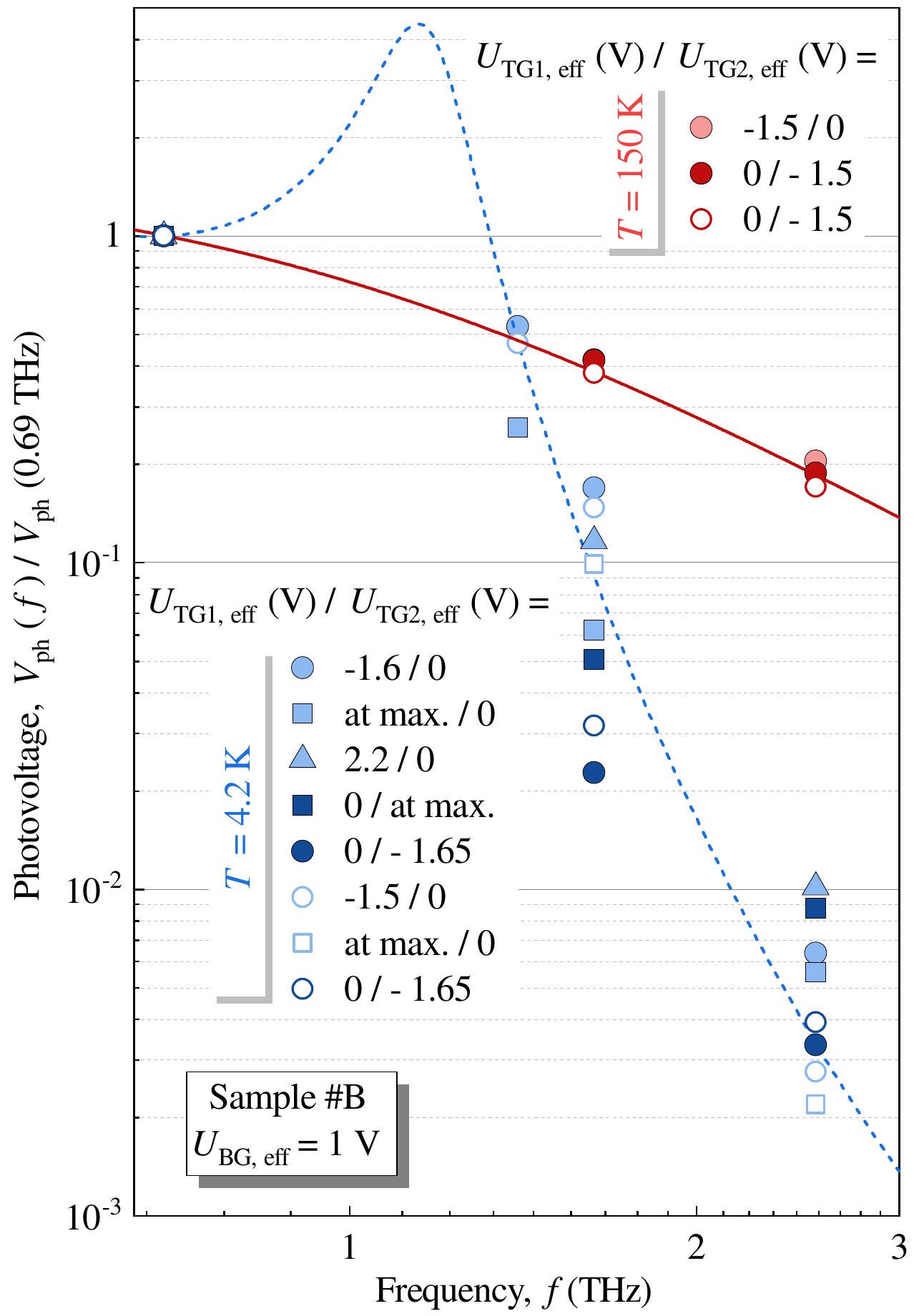}
	\caption{\sdg{Frequency dependencies \sdg{of both ratchet contributions, $V_0$ (filled symbols) and $V_{\rm L1}$ (empty symbols),}
	\sdg{measured in sample \#B for} various \sdg{sequences of top gate voltages ($U_{\rm TG1, eff}$ or $U_{\rm TG2, eff}$) and} 
	$U_{\rm BG, eff} = 1$~V. The data obtained at different temperatures are shown in red for $T = 150$~K and in blue for $T = 4.2$~K.
	Similar to Fig.~\ref{FigR6}(a) the red solid curve corresponds to Eq.~(\ref{Drude}) with $\tau = 0.17$~ps used to fit the data at $T = 150$~K, whereas the dashed blue curve represents the crossover from DD to HD following from Eq.~(\ref{final-current-large-s}). For better visualization a double logarithmic scale was used to present the data.
	}}
	\label{FigR6_2}
\end{figure}

\sdg{To highlight the drastic difference in the frequency dependence of the ratchet contributions, $V_0$ and $V_{\rm L1}$, \sdg{as compared to the Lorentz-curve}
%
we combined in 
\sdg{Fig.~\ref{FigR6_2}} all data obtained for $U_{\rm BG, eff} = 1$~V and two temperatures (4.2 and 150~K). The data, extracted from Fig.~\ref{FigR6}, were obtained at a low carrier density and exhibit the strongest difference in the frequency behavior. The data reveal that for $f = 1.69$~THz at various experimental conditions the difference in magnitude,  as compared to the Lorentz-curve, ranges between 2.5 and 20 times, whereas for $f = 2.54$~THz it becomes much larger and varies between 20 and 100 times. The large variation of amplitudes measured at high frequencies for 4.2~K is attributed to very different sequences of top gate voltages. We see that although the ratchet effect is sensitive to the details of the density profile controlled by the top gates (as expected from the theory shown below), the general trend of \sdg{strengthening} of the frequency dependence is clearly observed. \sdg{A comparison with the developed theory, see Sec.~\ref{Discussion}, shows that such drastic frequency dependence is expected for the THz-ratchet photoresponse in the hydrodynamic regime. The corresponding calculated curve is plotted in Fig.~\ref{FigR6_2} as a blue dashed line, whereas the Lorentz-curve fitting the data at $T=150$~K is presented by the red solid curve.} 
The comparison between the experimental data and the 
 \sdg{calculated curves} 
 will be presented in Sec.\ref{Discussion}.}


\section{Theory}
We start with a brief review of state of the art.
The standard calculations  of
the ratchet current~\cite{Ivchenko2011,Olbrich2016,Nalitov2012} are performed in the drift-diffusion approximation where  both electron-electron (ee) and electron-phonon  interactions are ignored as compared to fast impurity scattering (although thermalization of the distribution function is implicitly assumed). Such a ratchet effect is referred sometimes to as electronic  ratchet. Actually, the effect of the ee-interaction is twofold and can be quite strong. First of all, as we mentioned,  sufficiently fast ee-collisions   can  drive the system  into the hydrodynamic regime. Secondly, ee-interaction   leads to plasmonic oscillations, so that a new frequency scale, the plasma frequency, $\omega_p(q)$ appears in the problem, where $q$ is the   inverse characteristic spatial scale of the system. For grating-gate structures with the period $L$ it is given by $q= 2\pi/L$.

The
ratchet effect is dramatically enhanced in the  vicinity of plasmonic resonances.
This plasmonic enhancement
can be excluded experimentally by using high excitation frequencies, $\omega \gg \omega_p(q).$  However, even in this case,  one should  choose the drift-diffusion  or hydrodynamic approximation depending on the relation  between the momentum relaxation and the  ee-collision rates. While hydrodynamics is usually applied to describe  plasmonic effects, photogalvanic ratchets are mostly treated within the drift-diffusion approximation. The detailed study of the ratchet effects for both regimes, including theoretical and experimental analysis of transition between them is absent so far.   Hence, interpretation of  experiments requires a subtle analysis of applicability of the approximations used. The theory presented below shows that both regimes are two limiting cases of the same problem and suggests parameters controlling the transitions between them. \sdg{The theory is developed for diffusive assumption, $ql \ll 1$. Ballistic effects are not significant in the experiment because  this inequality  was satisfied  even for the lowest temperature. In particular, the results shown in Fig.~6 below, correspond to an electron density about $2.5 \times 10^{11}$~cm$^{-2},$ and, respectively, to the mean free path not exceeding $0.4$~µm, which is much smaller than the modulation period.}

 \subsection{Model}
We consider the  motion of 2D  electrons with parabolic energy dispersion
in the static  periodic  potential
\be
\label{V}
U(x)   = U_0\cos qx, \qquad q=\frac{2\pi}{L},
\ee
which arises in the 2D gas due to presence of the grating gate structure with the period $L$.

The external field  of general polarization is described by phases $\alpha$ and $\theta$:
\be
E_x = {E_{0}} \cos \alpha \cos {\omega}t, ~~
E_y = E_{0}\sin \alpha \cos \left( {{\omega}t + \theta } \right),
\label{Eext}
\ee
which are connected to the standard Stokes parameters
as follows:


\be
\begin{aligned}
	& P_{\rm L1}=\cos{2\alpha},
\\
	& P_{\rm L2}=\sin{2\alpha} \, \cos \theta,\\
	&P_{\rm C}=\sin{2\alpha} \, \sin \theta.
	\label{Stockes}
\end{aligned}
\ee
%
%
Note that for the $\lambda$-half and $\lambda$-quarter experimental setup arrangements these parameters take the form given in Sec.~\ref{results}.

We also assume that the grating leads to modulation of the electric field with the depth $h,$
so that the field acting in the 2D channel,
has spatially modulated
amplitude.
The components of total field in the channel read
\begin{align}
\label{Ex}
E_x (x,t) & = [1+ h \cos(q x+\phi)]  E_0 \cos \alpha  \cos \omega t ,
\\
\label{Ey}
E_{y}(x,t) & = [1+ h \cos(q x+\phi)]  E_0 \sin \alpha  \cos (\omega t+\theta),
\end{align}
where $\phi$ is the phase which determines the asymmetry of the modulation.

We search the response of the  2D electron system to the above described perturbation
by using two approaches -- hydrodynamic (HD)  and drift-diffusion (DD).
In both approaches,  the direction of the current is determined by the
frequency-independent
asymmetry parameter~\eqref{Xi}
\begin{equation}
\label{Xi_def}
\Xi=
 \frac{E_0^2 h U_0 q \sin \phi  }{4} ,
\end{equation}
which
is present due to
the phase shift   $\phi$ between the static potential and the near-field amplitude.

\subsection{Microscopic theory of limiting cases: drift-diffusion and hydrodynamic regimes}
\label{ZMF}

Hydrodynamic  theory formulates  equations  for the concentration  and velocity.
Calculations within the hydrodynamic approach  yield  (here, we rewrite  results of  Ref.~\cite{Rozhansky2015} in terms of the radiation Stokes parameters):
\BEA \label{j-hyd-x}
&& j_{x}^{\rm HD}=   \Xi ~\frac{e^3 N_0 \tau^3 q^2}{ m^3}  \frac{ 1 +P_{\rm L1}   }{(1 + \omega ^2 \tau^2)[(\omega ^2 - \omega _q^2)^2 \tau^2 +\omega ^2   ]  }, \nonumber
\\ \label{j-hyd-y}
&&j_{y}^{\rm HD} = - \Xi ~\frac{e^3 N_0 \tau^3 }{ 2 m^3 s^2} ~ \frac { (\omega ^2 - \omega _q^2)  P_{\rm L2}
	+ \frac{\omega}{\tau} P_{\rm C} }{(\omega ^2 - \omega _q^2)^2 \tau^2 +\omega ^2 }.
\EEA
Here  $\omega_q=s q$ is the plasma frequency,
$s$ is the plasma wave velocity,
and $\tau$ is the electron momentum relaxation time.

In the drift-diffusion theory, the ratchet current is obtained by sequential
iterations of the kinetic equation in the  weak electric field radiation
amplitude and the static force of the ratchet potential $\propto
\textrm{d}U/\textrm{d}x$. It was also assumed that the wave vector $q$ is sufficiently  small, so that one can keep linear in
$q$ terms only. Since the parameter $\Xi$ already contains factor $q$, see Eq.~\eqref{Xi_def},  it can be put
zero in all other terms. Also, the drift-diffusion approximation fully
neglects the electron-electron interaction.
Within such approximations, the components of the ratchet current
for parabolic energy dispersion are given
by~\cite{Ivchenko2011,Olbrich2016,Nalitov2012}
\begin{align}
&j^{\rm DD}_x=-\Xi ~
\frac{e^3 N_0 \tau^3}{m^3 v_{\rm F}^2}~
{1+ P_{\rm L1} \over 1+(\omega\tau)^2},
\label{j-dr-x}
\\
&j_y^{ \rm DD}=-\Xi ~
\frac{e^3 N_0 \tau^3}{m^3 v_{\rm F}^2}~
{P_{\rm L2} + {1\over \omega\tau}P_{\rm C} \over 1+(\omega\tau)^2}. \label{j-dr-y}
\end{align}
Here we do not take into account the so-called Seebeck ratchet contribution~\cite{Ivchenko2011,Olbrich2016,Nalitov2012} governed by the energy relaxation processes.

\subsection{Comparison of drift-diffusion and hydrodynamic regimes}
\label{comparisonHDDD}

Let us now compare results obtained within hydrodynamic and drift-diffusion approximations.
To this end, we first rewrite formulas for $j_{x,y}^{\rm HD} $ in the limit
of the sufficiently small $q,$ which was used in derivation of drift diffusion equations. We get
\BEA \label{j-hyd-x1}
&& j_{x}^{\rm HD} \approx
\Xi ~\frac{e^3 N_0 \tau^3 q^2}{ m^3 \omega^2}  \frac{ 1 +P_{\rm L1}  }{(1 + (\omega\tau)^2)^2  },
\\ \label{j-hyd-y1}
&&j_{y}^{\rm HD} \approx - \Xi ~\frac{e^3 N_0 \tau^3 }{ 2 m^3 s_0^2} ~ \frac { P_{\rm L2}
	+ \frac{1 }{\omega\tau} P_{\rm C}}{1+ (\omega\tau)^2},
\EEA
for $q \to 0.$ Comparing Eqs.~\eqref{j-hyd-x1} and \eqref{j-hyd-y1} with Eqs.~\eqref{j-dr-x}
and \eqref{j-dr-y} we see that $j_{y}^{\rm HD}= j_{y}^{\rm DD}.$ Indeed,  the plasma wave velocity
is given by
\be
s_0 =  \sqrt{\frac{e^2 N_0}{m C} +\frac{v_{\rm F}^2}{2}},
\label{s}
\ee
where $ C= \epsilon /4 \pi D_s $
is the channel  capacitance per unit area, $D_s$ is the distance to the spacer,  $\epsilon $
is the dielectric constant,
$N_0$ the concentration in the channel,
 and the term $v_{\rm F}^2/2$    represents   the contribution of the
Fermi pressure.    In the absence of the interaction, the latter  contribution dominates.  Replacing in
Eq.~\eqref{j-hyd-y1},  $s_0^2 \to v_{\rm F}^2/2,$ we reproduce Eq.~\eqref{j-dr-y}. On the other hand,
although  polarization dependences of   $x$-components of the current is the same, they  are parametrically different and, moreover, have different signs
\be
\frac{j_x^{\rm HD}}{j_x^{\rm DD}}= - \frac{v_F^2 q^2}{\omega^2 (1+ \omega^2 \tau^2) }.
\label{criterii} \ee
The most important difference, which can be checked experimentally, is  the      frequency
dependence  difference both at high  and low frequencies. For  high frequencies
hydrodynamic  current decays much faster:
\be
j_x^{\rm HD} \propto \frac{1}{\omega^ 6},\quad   j_x^{\rm DD} \propto  \frac{1}{\omega^2},
\qquad \text{for}\quad \omega \to  \infty.
\ee
For low  frequency, $\omega \ll 1/\tau ,$  $j_x^{\rm HD}$ diverges while  $j_x^{\rm DD}$ saturates
\be
j_x^{ \rm DD} \to  \text{const},\quad \text{for}\quad \omega \ll 1/\tau.
\ee
We also notice
that $y$-components of the current calculated in two different approximations coincide only in the
limit $q \to 0.$ For any finite $q$ there is an essential difference.
The circular contribution diverges at $\omega \to 0$ in both approaches.
In the hydrodynamic regime this divergence is cured
by  the Maxwell relaxation.
In the drift-diffusion approximation, it
is cured by the energy relaxation caused by
the electron-electron or electron-phonon interaction~\cite{Nalitov2012}.
One can expect that inclusion
of  the inelastic scattering into hydrodynamic approach would  lead to restriction of response
by at the frequencies of the order of $1/\tau_{\rm M} + 1/\tau_{\rm ee}.$
Also, taking into account finite $q$ in the drift-diffusion model would limit the divergence of
the response at low frequency by conventional diffusion at
$\omega \sim D/L^2$ (here $D$ is the diffusion coefficient) which replaces the Maxwell relaxation rate.

Above, we discussed the non-resonant case $\omega \gg \omega_q$.
It makes little sense to compare the two approximations in the
opposite resonant regime,
since the drift-diffusion approximation  as a starting point  assumes
that $ q $ is small and, as a consequence, the  inequality $\omega \gg \omega_q$
is fulfilled.
For not too small $q,$ resonant conditions,
\be
\omega_q \tau \gg 1, \quad \omega-\omega_q \sim 1/\tau,
\ee
can be satisfied and plasmonic resonances appear in the ratchet effect~\cite{Rozhansky2015}.
One might expect that analogous resonances would appear in the drift-diffusion
approximation provided higher order in-$q$ terms are taken into account. This would happen even
in the absence of the electron-electron interaction due to the Fermi sound [see second term in
the square root in   Eq.~\eqref{s}].

\subsection{Transition from the hydrodynamic to the drift-diffusion regime}
\label{TransitionHDDD}

In this section, we develop a theory  describing transition from HD to DD approximation.  We limit ourselves by the case of linear  polarization, parallel to $x$ axis.  As a key assumption, which essentially simplifies calculations  and allows us to get an analytical expression for dc current, which shows HD-DD crossover,    we use  a model form of the electron-electron collision  integral. Also, we do not discuss here effect of the electronic viscosity, as well as effects related to heating of the
electron gas.

We start with the kinetic equation
\be
\frac{\partial f}{\partial t} + \bm{v} \cdot \frac{\partial f}{\partial \bm{r}}+\frac{\bm{F}}{m} \cdot \frac{\partial f}{\partial \bm{v}}=\frac{\langle f \rangle_{\theta}-f}{\tau}+\frac{f_{\rm{HD}}-f}{\tau_{\rm{ee}}},
\ee
where $m$ is the effective mass,  ${(\langle f \rangle_{\theta}-f)}/{\tau}$ is impurity collision  integral ($\langle \cdots \rangle_\theta$ stands for velocity angle averaging) and
$\bm F$ includes both external field and the plasmonic field  induced by the inhomogeneous electron concentration.
We use the model form of the    electron-electron collision integral~\cite{Gross1956}
${(f_{\rm{HD}}-f)/\tau_{\rm{ee}}}$,
where
$f_{\rm{HD}}$ is  hydrodynamic function having standard form
\be
f_{\rm{HD}}(t,\mathbf{r},\mathbf{v})=\frac{1}{\exp\left[\frac{m\left(\mathbf{v}
-\bm{V(\mathbf{r},t)}\right)^2/2-\mu(\mathbf{r},t)}{T(\mathbf{\mathbf{r},t})}\right]+1},
\label{fHD}
\ee
with local equilibrium  parameters   $\bm V(\mathbf{r},t),$ $   \mu(\mathbf{r},t),$ and  $T(\mathbf{\mathbf{r},t}). $  These parameters are found by using unknown function $f,$ which, in general case, does not have hydrodynamic form. Evidently,
\be
\mu= N/g,~N=\int f \{dp\}, \quad \bm V=\frac{1}{N}\int \m v f   \{dp\},
\label{mu-V}
\ee
where $N$ is the electron concentration, $g$ is the density of states (with account for spin and valley degeneracy)
and $\{dp\}={d^2 \mathbf{p}}/{(2 \pi \hbar)^2}.$ In order to find expression of  $T$ via $f,$ one needs to multiply  $(f_{\rm{HD}}-f)/\tau_{\rm{ee}}$  by $\epsilon = m v^2/2$ and integrate over  $\{dp\}$ having in mind that the ee-collisions conserve the electron energy. We thus obtain
\be
\int v^2 f \{dp\}=  \frac{2g}{m}\left(\frac{\mu^2}{2}+\frac{\pi^2 T^2}{6}\right)+V^2 g \mu.
\label{balance}
\ee
This equation defines the expression of the  electronic temperature in terms of function $f$ with $\mu$ and $\bm V$ found from  Eq.~\eqref{mu-V}.  Physically, Eq.~\eqref{balance} follows from the   energy conservation  law.

Multiplying now kinetic equation by  $1, ~\m v,~ v_i v_k$ and integrating over momenta, we get the following equations
\begin{align}
\frac{\partial N}{\partial t}& +\nabla_{i}J_{i}=0,
\label{N}
\\
\frac{\partial J_{\rm{i}}}{\partial t}& +\nabla_{\rm{k}}J_{\rm{ik}}-\frac{F_{\rm{i}}}{m} N=-\frac{J_{\rm{i}}}{\tau},
\label{J}
\\
\label{Jik}
\frac{\partial J_{\rm{im}}}{\partial t}&
-\frac{1}{m}\left(F_{\rm{i}} J_{\rm{m}}+F_{\rm m} J_{\rm i}\right)
\\
&=\left(\frac{1}{\tau_{\rm ee}}+\frac{1}{\tau}\right)\left(\frac{\delta_{\rm im}}{2}{\rm Tr}[J_{\rm im}]-J_{\rm im}\right)\nonumber
\\
&+\frac{N}{\tau_{\rm ee}}\left(V_{\rm i} V_{\rm m}-\delta_{\rm im}\frac{V^2}{2}\right). \nonumber
\end{align}
Here,
\be
\m J= \int  \m v f \{dp\}, \qquad J_{\rm ik}=
\int   v_i v_k f \{dp\}.
\ee
Equations \eqref{N} and \eqref{J} represent, respectively, the continuity equation   and  Euler-like equation.   Equation~\eqref{Jik} needs some clarifications. First of all, we skipped  in this equation the term $\p_k \int v_i v_m v_k f \{dp\} ,$ responsible for heat convection.  Secondly, we note that trace of
this    equation     gives  infinite heating  because we neglected cooling by phonons.    Actually, this cooling exists and limits the electron temperature.   Then one can use the following method for solution of
Eqs. \eqref{N}--\eqref{Jik}.
First, we write
$$J_{\rm ik}=(\delta_{\rm ik}/2) \rm Tr \hat J +\Lambda_{ ik},$$
where ${\rm Tr} \hat \Lambda=0.$
Equation \eqref{Jik}  can be used for finding  traceless part  $\Lambda_{\rm ik}.$  As for   ${\rm Tr} J_{\rm ik},$   it is determined  by  Eq.~\eqref{balance}. Finally, we neglect term  $\propto T^2$ in this equation assuming that cooling by  phonons limits temperature on a sufficiently low level $T\ll \mu .$
Then,  we arrive at the following system of equations

\begin{align}
& \frac{\partial N}{\partial t}+\nabla_{\rm i}J_{\rm i}=0, \\
& \frac{\partial J_{\rm i}}{\partial t}+J_{\rm i}\gamma+\nabla_{\rm i}\left(\frac{V^2 N}{2}+\frac{\mu^2 g}{2 m}\right) + \nabla_{\rm k}\Lambda_{\rm ik}=f_{\rm i} N, \\
& \left(\frac{\partial }{\partial t}+\gamma+\gamma_{\rm ee}\right) \Lambda_{\rm ik}=\left(f_{\rm i} J_{\rm k}+f_{\rm k} J_{\rm i}\right)-\delta_{\rm ik}
\m f \cdot \m J \\
&+\gamma_{\rm ee} N\left(V_{\rm i} V_{\rm k}
-\delta_{\rm ik}\frac{V^2}{2}\right). \nonumber
\end{align}
Here $f_{\rm i}=F_{\rm i}/m$, $\gamma=1/\tau$, and $\gamma_{\rm ee}=1/\tau_{\rm ee}$.
Next, we assume that local current flows everywhere  in $x$-direction and that   all variables depend on the $x$ coordinate only:
$\bm{J}\parallel\bm{e_{\rm x}}$, $\Lambda_{\rm xx}=-\Lambda_{\rm yy}=\Pi.$
We skip index $x$ below ($J_{\rm x}=J,~ V_{\rm x }=V,~ f_{\rm x}=f$)
thus arriving to the following set of coupled equations
\be
\begin{aligned}
& \frac{\partial N}{\partial t}+\nabla_{\rm x} J=0, \\
& \frac{\partial J}{\partial t}+J\gamma+\nabla_{\rm x}\left(\frac{V^2 N}{2}+\frac{\mu^2 g}{2 m}\right) + \nabla_{\rm x}\Pi=f  N ,\\
& \frac{\partial \Pi}{\partial t}+(\gamma+\gamma_{\rm ee})\Pi=f J+\gamma_{\rm ee} \frac{N V^2}{2}.
\end{aligned}
\label{system0}
\ee
Solution of these equations can be found in a full analogy with the hydrodynamic
case (see Ref.~\cite{Rozhansky2015}).  We delegate the calculations to Appendix~\ref{App_theory}.  For arbitrary relation between  $s$ and $v_{\rm F},$  the expression for the current is quite cumbersome and is given by Eqs.~\eqref{Jdc-(tot)}, \eqref{j-final-n11}, and \eqref{j-final-n20}.    This
 expression  simplifies under the assumption $s \gg v_{\rm F} $:
\be
\frac{J_{\rm dc,x}}{C_1}=    \frac{\omega^2(\gamma^2+\omega^2)+\omega_{q}^2[2\gamma(\gamma+\gamma_{\rm ee})-\omega^2]}{(\gamma+\gamma_{\rm ee})(\gamma^2+\omega^2)[\gamma^2\omega^2+(\omega^2-\omega_{q}^2)^2]} .
 \label{final-current-large-s}
 \ee
Here $\omega_q = s_0 q$
 is the plasma wave frequency with account of the sound contribution, cf. Eq.~\eqref{s},
 and
\be
C_1
={e^3 N_0\over  2 m^3 s_0^2}~\Xi
\label{C0}
\ee
with $\Xi$ given by Eq.~\eqref{Xi}. In derivation of Eq.~\eqref{final-current-large-s} we assumed $s_0\approx s.$

In the limit  $\gamma_{\rm ee}\to\infty$ we reproduce
HD equations:
\be
\frac{J_{\rm dc,x}}{C_1}=\frac{ 2 \tau \omega_{q}^2 }{(1+\tau^2\omega^2)[\omega^2/\tau^2 + (\omega_{q}^2 -\omega^2)^2]}.
\ee
In the opposite limit $\gamma_{\rm ee}\to 0$  we arrive at the DD equation
\be
\frac{J_{\rm dc,x}}{C_1}=\frac{   \tau^3}{1+\tau^2\omega^2}
\left\{ 1+\frac{\omega_q^2[2+ \tau^2(\omega^2-\omega_q^2)]}{  \omega^2 + \tau^2(\omega^2-\omega_q^2)^2} \right\}.
\ee
This  equation is further simplified in the limit $q\to 0$ to
\be
\frac{J_{\rm dc,x}}{C_1}=\frac{  \tau^3}{1+\tau^2\omega^2}.
\ee

\begin{figure}[h!]
	\centering
	\includegraphics[angle=0,width=0.95\linewidth]{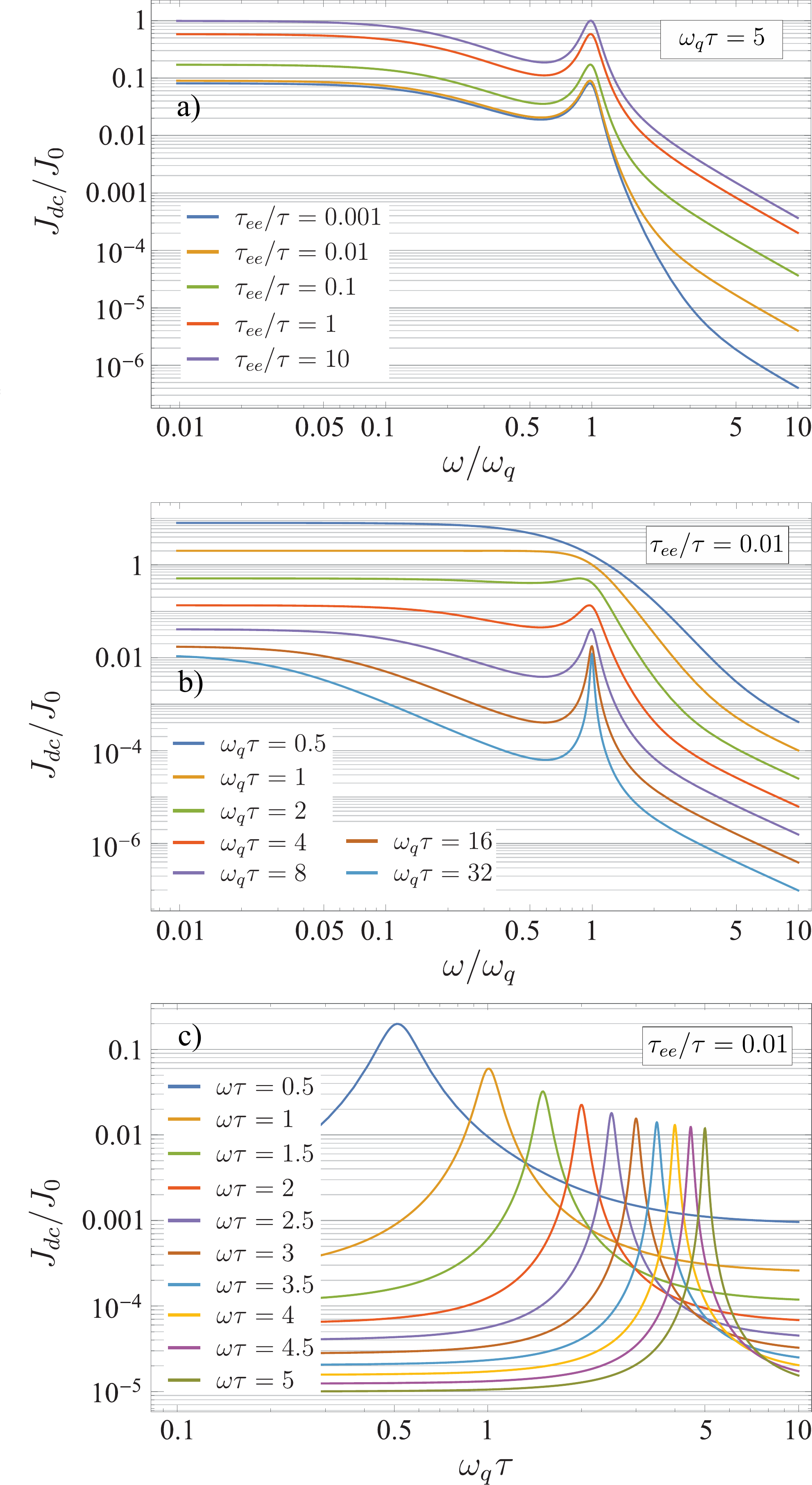}
	\caption{Theoretical dependencies following from  Eq.~\eqref{final-current-large-s}:  (a) Dependence of current on  $\omega/ \omega_{q}$, for $\omega_{q}\tau=5$ and different values of $\tau_{\rm ee}/ \tau$ covering both HD and DD regimes, (b) Frequency dependence of the dc response for different quality factors, (c) Dependence of the response on the quality factor for fixed different $\omega.$}
	\label{FigT}
\end{figure}

 The frequency dependencies of $J_{\rm dc,x}$ given by Eq.~\eqref{final-current-large-s} are analyzed  in Fig.~\ref{FigT}.  In  Fig.~\ref{FigT}(a),
we plot the dependence of the  dc response on the frequency  for fixed quality factor, $\omega_q \tau=5$, and different values of $\tau_{\rm ee}/\tau$  ranging from
0.001 (HD regime)  to 10 (DD regime).  We see that the response
shows the plasmonic resonance both in HD and DD regimes. We also  see that with
decreasing   $\tau_{\rm ee}/\tau$ the high-frequency behavior evolves from
slow  $1/\omega^2$  to  much sharper $1/\omega^6$ law.  In  Fig.~\ref{FigT}(b) we plotted the response  as a function of frequency for the small value of $\tau_{\rm ee}/\tau=0.01$ corresponding to  HD regime,    for different quality
factors. We see that, as expected, the plasmonic resonance becomes sharper with
increasing $\omega_q \tau.$  At the same time,  the high-frequency asymptotic
of the  current is not sensitive to the quality factor and  is given by
$1/\omega^6$ dependence.  In   Fig.~\ref{FigT}(c),  we plotted  the dc current  as a function  of $\omega_q\tau$ for  fixed different values of $\omega.$ We see, that  the resonance
occurs at  $\omega_q\approx \omega.$

%

\section{Discussion}
\label{Discussion}

\begin{figure}[h]
	\center{\includegraphics[width=0.95\linewidth]{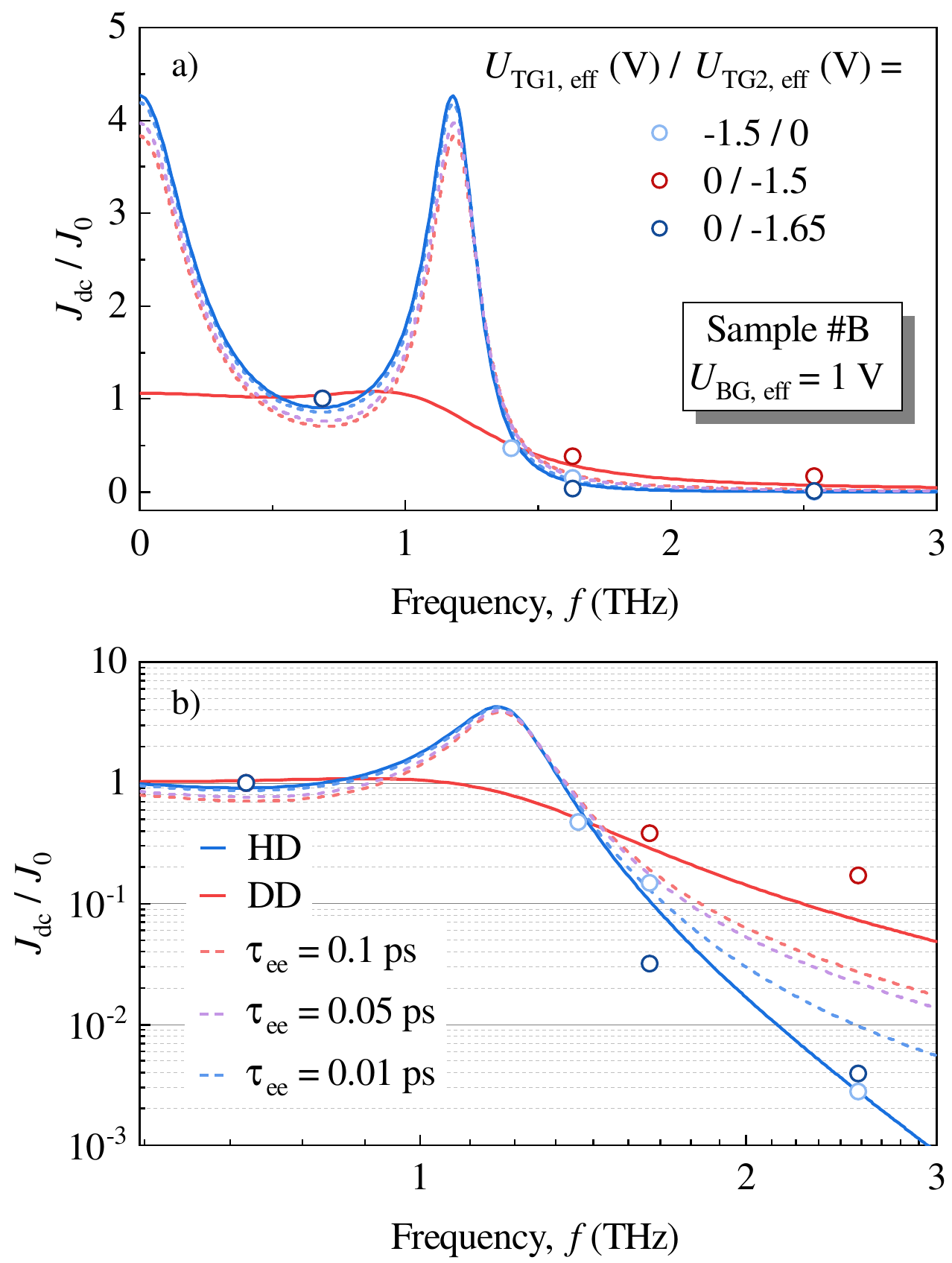}}
	\caption{Frequency dependence of the photocurrent, $J_{\rm dc}$, according to
		Eq.~(\ref{final-current-large-s}) showing the crossover from DD to HD regime with increasing $\gamma_{\rm ee}/\gamma$.  
			\sdg{Panel a) shows the experimental data for the linear ratchet contribution ($V_{\rm L1}$) and the corresponding theoretical curves presented in a double linear scale, panel (a), and in double logarithmic scale, panel (b). The latter is used because for $f = 1.63$ and 2.54~THz the signals differ up to two orders of magnitude.}
		The theoretical curves correspond to the following parameters: $\tau = 0.1$~ps, $\tau_{\rm ee}=900$~ps, $f_{q}= 1.16$~THz \sdg{(red curve)} 
		and $\tau = 0.7$~ps, $\tau_{\rm ee}=0.001$~ps, $f_{q}= 1.16$~THz (blue curve). \sdg{The dashed curves were calculated by using $\tau = 0.7$~ps and a corresponding $\tau_{\rm ee}$ as labeled in the figure.}}
	\label{ris:image1}
\end{figure}

Above we presented the theoretical curves
for different values of parameters (see Fig.~\ref{FigT}). Let us now compare the developed theory and experimental data. 
\sdg{For that we present in Fig.~\ref{ris:image1} the results obtained for the case of linear ratchet contribution, $V_{\rm L1}$, high enough top gate voltages and two temperatures. The choice is because our theory is developed for such conditions.}
Although the experimental precision did not allow us to identify plasmonic resonances, we were able to distinguish  between different high-frequency asymptotic dependencies specific for HD and DD regimes.

Our experimental results on three graphene structures with different parameters of DGG provide a self-consistent picture demonstrating that the photosignals (photocurrents) are generated due to the presence of asymmetric superlattices and consequently controllable lateral asymmetry parameter $\Xi$.  The  change of sign upon reversing the in-plane asymmetry of the electrostatic potential as well as changing the carrier type clearly demonstrate that they are caused by the ratchet effect. Corresponding experimentally results shown in Figs.~\ref{FigR1} and~\ref{FigR4},  inset in Fig.~\ref{FigR5}(a), and  Fig.~\ref{FigR2} in Appendix~\ref{Additional_measurements}, are  in full agreement with theoretical Eqs.~\eqref{j-hyd-x}-\eqref{j-dr-y}.  Observed photocurrents 	are characterized by specific polarization dependencies revealing substantial contributions of all three ratchet effects: polarization-independent, linear and circular one, for polarization dependencies see experimental Fig.~\ref{FigR1} and Eqs.~\eqref{j-hyd-x}-\eqref{j-dr-y}. Comparing  the magnitudes of the ratchet effect detected in bilayer graphene DGG structures (current work) with those in monolayer graphene~\cite{Hubmann2020} we obtained that they are close to each other (same order of magnitude, more precise comparison can not be done because e.g. the DGGs in these samples are not the same). Alike reported previously for the monolayer graphene, we also detected that at low temperatures the ratchet effects are enhanced in the vicinity of the Dirac point, see Fig.~\ref{FigR3} Appendix~\ref{Additional_measurements}.
	
Most remarkably,  we identified  several  regimes, where the frequency dependence of signal is qualitatively different, see Fig.~\ref{FigR6}. The most substantial difference is observed between frequency dependencies of the ratchet effect excited in samples at high temperature (150~K), $j \propto 1/\omega^2$,  and liquid helium temperature ($j \propto 1/\omega^6$, for low carrier density controlled by the back gate voltage), see Fig.~\ref{FigR6}(b).   We demonstrate experimentally  crossover  between these regimes  with changing temperature, see Fig.~\ref{FigR6}(b), and concentration, see Fig.~\ref{FigR6}(c). We attribute this remarcable result to the transition from drift-diffusion to hydrodynamic regime, see Sec.~\ref{TransitionHDDD}. We developed a theory which allows one to describe a transition from the DD to the HD regime.   In particular, Eq.~\eqref{final-current-large-s} 	reveals dependence on $\gamma_{\rm ee}$ and $\gamma$ and shows $ J_{\textrm{dc},x} \propto 1/\omega^2$ high frequency  asymptotic  for $\gamma_{\rm ee} \to 0$ and  $J_{\textrm{dc},x} \propto 1/\omega^6$  for  $\gamma_{\rm ee} \to \infty.$  We demonstrate that Eq.~\eqref{final-current-large-s}   describes the experimental data very well if one uses $\gamma$ and $\gamma_{\rm ee}$ as fitting parameters, as seen from Fig.~\ref{ris:image1}. We see that the high-temperature ($T=150$~K) data are well fitted by DD model. Physically, this is clear  because at such high temperature transport is fully dominated by phonons. This is not the case for  low temperature, $T=4.2$~K.  For such a temperature, crossover to the HD regime with much stronger $\omega$ dependence is clearly seen in the experiment. Let us discuss this point in detail.  Our experimental results on frequency dependence are shown in Fig.~\ref{FigR6}, where data for three different temperatures: $T=2$, 4.2, and $150$~K  as well as for different  gate voltages are presented in double logarithmic scale.  As it is seen, data obtained at $T=150$ K are perfectly fitted by the Lorentz-curve with the asymptotic behavior $1/\omega^2$. We also see that for a relatively low temperature, $T=4.2$~K, the high-temperature dependence follows $1/\omega^6$, which is in a good agreement with the HD model. On the other hand, further decrease of the temperature to $T=2$~K moves the system back to the DD regime. Such a non-monotonous dependence can be easily understood.  Indeed,  for high temperatures, the phonon scattering dominates over $\gamma$ and $\gamma_{\rm ee}$. For a temperature $T=2$ to $4.2$~K the phonon scattering is negligible, so that we have a competition between $\gamma$ and $\gamma_{\rm ee}$. The rate of the electron-electron scattering increases with the temperature and decreases with the electron concentration:
\be \gamma_{\rm ee} \propto T^2/E_F.
\label{gamma-ee}
\ee
(Here, we limit ourselves by a very rough estimation. For discussion of more complicated models  see recent publication~\cite{Alekseev2020} and references therein). Hence at a very low temperature, the electron-electron collisions are also negligible and we return to the DD regime. This is clearly seen in Fig.~\ref{FigR6}(b): with lowering $T$ to 2~K the frequency dependence becomes less sharp.  The dependence  on back gate voltage
is also consistent with Eq.~\eqref{gamma-ee}. Indeed, as seen from Fig.~\ref{FigR6}(c) the data for $U_{\rm BG,eff}=1$~V are well fitted by $1/\omega^6$, while the data for  $U_{\rm BG, eff}=5.5$~V  are closer to a $1/\omega^2$-dependence.

The direct comparison of the theoretical formula Eq.~\eqref{final-current-large-s} with the experimental data is presented in Fig.~\ref{ris:image1}. We see that the experimental data for $T=4.2$~K can be well fitted by Eq.~\eqref{final-current-large-s} with $\gamma_{\rm ee} = (50-70) \gamma$~\footnote{Note that the values of $\tau_{\rm ee}$ in unstructured graphene at low temperature, which we found in the literature are of the order of 0.1-1~ps~\cite{Kumar2017,Drienovsky2018,Ho2018}.}.
\newline

\sdg{Finally we note, that while so far we discussed only the data corresponding to the conditions satisfying those considered in the theory (linear polarization and the gate voltages far from the charge neutrality point) the strong frequency dependence has also been observed for polarization-independent ratchet contribution and gate voltages in the vicinity of the charge neutrality point. These experimental results are summarized in Fig.~\ref{FigR6_2} and demonstrate that at $T = 4.2$~K and low carrier density the ratchet currents closely follows the calculated curve (blue dashed curve in Fig.~\ref{FigR6_2}, which is similar to the solid blue curve in Fig.~\ref{ris:image1}).}
	\newline

Let us 
\sdg{summarize} our key findings:
	\begin{itemize}
	\item We demonstrated experimentally  strong ratchet effect in bilayer graphene and find its frequency dependence
\item At high temperatures ($T \geq 150$~K), the dc photoresponse  varies with the radiation frequency  according the Lorentz formula with $1/\omega^2$ asymptotic behavior  in a good agreement with the  drift-diffusion model.
	\item Strikingly at liquid helium temperature the response  exhibits drastic frequency dependence with  $1/\omega^6$ asymptotic  behavior, which is perfectly fitted  by HD equation
	
	\item Variation of temperature in the interval $T=2-4.2$~K  and concentration (by back gate potential)    show  crossover between $1/\omega^2$ and $1/\omega^6.$
	
	\item{We derived analytical expression which captures basic experimental results. In particular,  it  describes crossover between drift-diffusion and hydrodynamic regimes. }
	
\end{itemize}

 \sdg{A few comments need to be made on the temperature dependence of the effect and on    the differences   of  our approach with previous analysis of viscous  flow   of the electron liquid.
The temperature range in which the observation of the hydrodynamic regime is possible is limited in both from above and from below. The lower limitation is caused by the decrease of the electron-electron scattering rate with the temperature, whereas the scattering rate on impurities is almost temperature-independent. 
Therefore, at sufficiently low temperatures, electron-electron scattering is "turned off" and the system goes into the drift-diffusion regime. On the other hand, with increasing temperature, sooner or later, strong scattering by phonons comes into play, which in the context of the problem under study is like scattering by impurities. Accordingly, there is an upper limit on the temperature. 
Importantly, as one can see from our data, the lower temperature limit for realization of the HD regime in our case is smaller as compared to realization of the viscous flow of the electron fluid studied experimentally, e.g. see in Refs.~\cite{Jong1995, Bandurin2016}. Indeed, one of the hallmarks of the viscous Poiseuille flow is the Gurzhi effect predicted in Refs.~\cite{Gurzhi1963, Gurzhi1965, Gurzhi1968}. Starting from its first experimental observation in Ref.~\cite{Jong1995},  this effect is considered as one of the most convincing arguments in favor of viscous transport. The Gurzhi effect is observed in a system of finite transverse (with respect to electron flow)  width $d$ under the assumptions $l_{\rm ee} \ll d \ll L_G,$ where $l_{\rm ee}$ is the electron-electron collision length, $L_G =\sqrt{l l_{\rm ee}}$ is the so called Gurzhi length, and $l$ is the mean free path with respect to impurity scattering. The inequality $d \ll L_G $ is not easy to satisfy in a sufficiently wide sample. That is why for the observation of viscous transport one have to use narrow-channel samples with ultrahigh mobility. Also, for the observation of negative non-local resistance, see Ref.~\cite{Bandurin2016}, the size of the viscosity-induced whirlpools responsible for viscous back flow (this size is in the order of $L_G$) was in the order of the size between contacts probing a negative voltage drop. If one uses thin wires or narrow strips for the observation of the Gurzhi effect, the second inequality, $l_{\rm ee} \ll d$ can be satisfied only at sufficiently high temperatures. By contrast, we study a bulk effect which does not disappear with increasing system size and distance between contacts. For the observation of the HD transport, we only need the condition $l_{\rm ee} \ll l,$ which is independent of the system size.}

\sdg{Moreover, the HD regime is not necessarily viscous. The key property of the HD regime (as compared to the DD one) is the presence of only three collective variables (local temperature, concentration and drift velocity), which completely characterize the system, in contrast to the DD regime, where the distribution function is not reduced to a hydrodynamic ansatz depending, in the general case, on an infinite number of variables. This regime can be realized for the case when the viscous contribution to the resistivity is small. This gives an additional possibility to observe the HD regime being exactly the case, which we studied in the current work. Importantly, we studied the non-linear regime with respect to the exciting field (which contains both static and dynamical parts and the final result is obtained only in the third order with respect to the total field). In such a regime, in contrast to the linear one, the difference between the distribution functions in the DD and the HD regimes is very strong and causes currents, which are strongly different even if one neglects viscosity. This allows one to distinguish between the DD and the HD regimes under the condition $d \gg {\rm max} (L_G,l).$ }

\sdg{We note also that the search of the HD regime in the static transport measurements at low temperature is difficult not only due to the low rate of electron-electron scattering, insufficient to realize condition $l_{ee} \ll d$ but also due to the presence of quantum corrections (see discussion in Ref.~\cite{Bandurin2016}) which can also contribute to a negative resistance considered as the main evidence of the HD regime in Ref.~\cite{Bandurin2016}. That is why Ref.~\cite{Bandurin2016}  is focused on the study of temperatures higher than 20 K. In our case, quantum corrections are suppressed because of the high-frequency measurements.}

\sdg{The above discussion can qualitatively explain why we observe the HD regime for lower temperatures as compared to Refs.~\cite{Jong1995,Bandurin2016,Crossno2016}. A more detailed study requires microscopic calculations of all scattering times involved in the problem (momentum relaxation time due to scattering by impurities, electron-electron collision rate, and electron-phonon scattering rate) and including 
the differences between parabolic and linear spectra. In this context, we notice that the purpose of this work is not to develop a quantitative approach to the problem and to carry out detailed microscopic calculations, which make it possible to determine the temperature boundaries of the HD range. Such research is beyond the scope of this work, in which we have developed a phenomenological approach to the problem.
We  demonstrated that even  this  simplified, involving a model form of the  electron-electron collision integral, allows one to qualitatively explain the change in the frequency dependence observed in our experiments and the tremendous  suppression of the high-frequency response in comparison with the generally accepted predictions within the framework of the DD model.}

\section{ Summary}
\label{summary}
To conclude, we observed and  studied  in detail the ratchet effect in bilayer graphene, focusing on the frequency dependence of the effect. We clearly identified two different regimes  with a qualitatively different frequency dependence corresponding to a     smoother and sharper frequency response. We also developed a theory that allows us to interpret these results as a transition from drift-diffusion regime  to hydrodynamic regime.

\sdg{To conclude the paper, we would like to stress two key points:} 

\sdg{(i) At present, the generally accepted model, within which photovoltaic phenomena in spatially modulated media are described, is the DD model, which predicts an $1/\omega^2$ dependence for the ratchet current. This frequency dependence was  studied  in  this work in detail. Our experimental results unequivocally demonstrate that the high-frequency response is not described by an $1/\omega^2$ dependence and for high frequencies is several orders of magnitude smaller. The observed high-frequency data are much better fitted by $1/\omega^6$ dependence. }

\sdg{(ii) The conclusion about HD-DD transition is based not only on our experimental results, but also, to a large extent, on theoretical calculations. We have developed for the first time a theoretical model that allows us to describe the transition from the DD to the HD regime. This model also unequivocally predicts fundamentally different frequency dependencies for these regimes. }

\sdg{Combination of the results (i) and (ii)  gives a very serious support of our central statement about DD-HD transition. }

\section*{Acknowledgments}


The support of  the DFG-RFFI project (Ga501/18, RFFI project 21-52-12015), IRAP  Programme  of  the Foundation   for   Polish Science   (grant   MAB/2018/9, project CENTERA), and the Volkswagen Stiftung Program (97738) is gratefully acknowledged.
J.E. acknowledge support of the DFG
through SFB 1277 (project-id 314695032, subproject A09).
%
The work of V.K. and S.P. was  also supported  by Russian Foundation for Basic Research grant No. 20-52-12019.
L.E.G. acknowledges the support of RFBR (project 19-02-00095) and Foundation for the Advancement of Theoretical Physics and Mathematics ``BASIS''.
K.W. and T.T. acknowledge support from the Elemental Strategy Initiativeconducted by the MEXT, Japan (Grant Number JPMXP0112101001) and JSPSKAKENHI (Grant Numbers 19H05790 and JP20H00354).
%

\appendix


\section{Transport and magnetotransport}
\label{Transport}

Electron transport and magnetotransport measurements were limited to the two-terminal method where the voltage drop was measured with a lock-in amplifier, while a low ac current $I_\mathrm{ac} = \MicroAmpere{0.1}$ at $f = \Hertz{12}$ was applied through $\SI{10}{\mega\ohm}$ series resistor.

\begin{figure}
	\centering
	\includegraphics[width=0.95\linewidth]{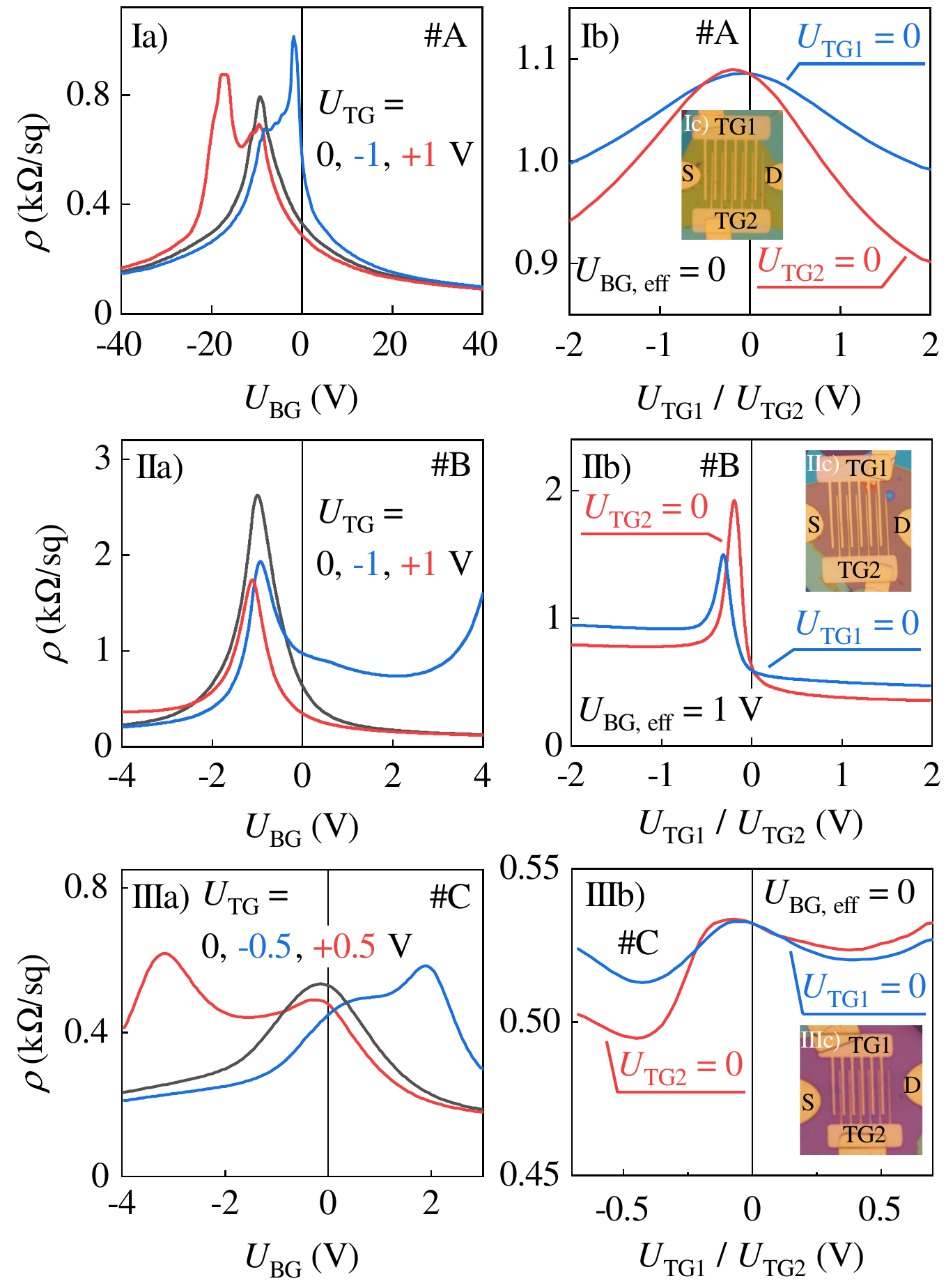}
	\caption{Two-terminal sheet resistivity $\rho$ as a function of back and top gate voltages at $T = 4.2$~K. The data are presented for samples \#A, \#B and \#C. $U_{\rm TG} = U_{\rm TG1} = U_{\rm TG2}$.
		Insets in panels (Ic), (IIc) and (IIIc) are optical micrographs of the samples.}
	\label{FigT1}
\end{figure}


\begin{figure}[h!]
	\centering
	\includegraphics[width=0.95\linewidth]{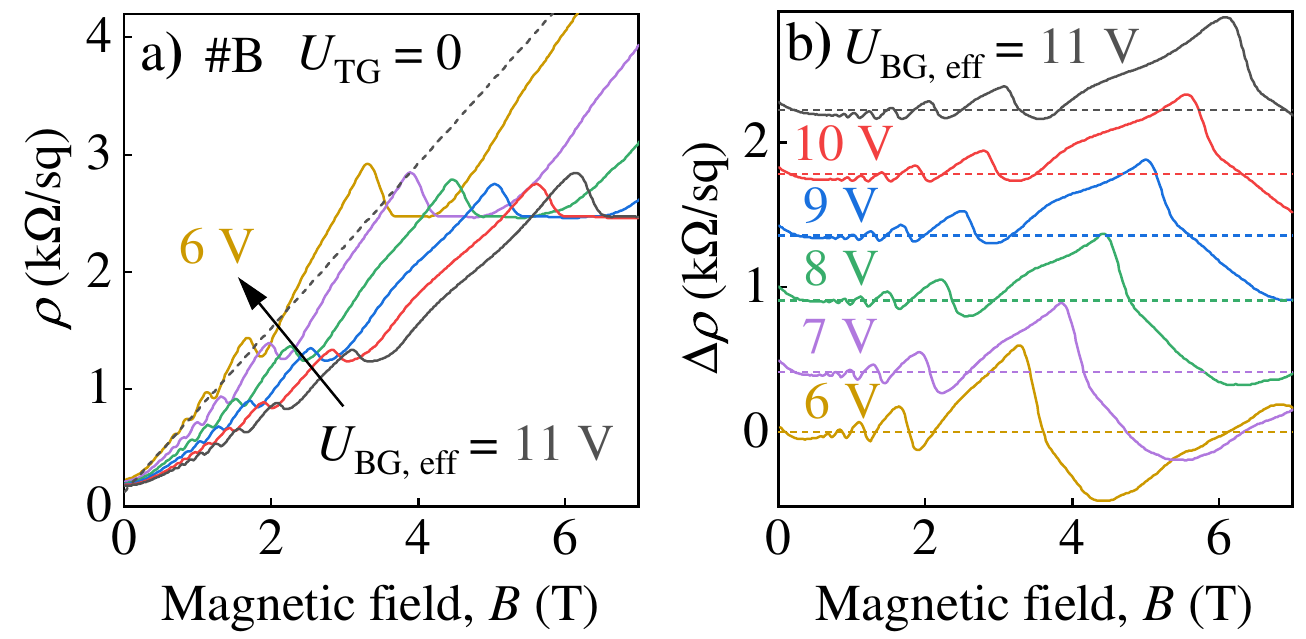}
	\caption{(a) Magnetic field dependence of two-terminal sheet resistivity $\rho$ measured in sample \#B at $T = 4.2$~K. (b) The oscillatory part $\Delta\rho$ was obtained by subtraction of the Hall contribution indicated by the dashed black line in panel (a). The data are presented for different effective back gate voltages and zero top gate bias, $U_{\rm TG} = U_{\rm TG1} = U_{\rm TG2} = 0$. The curves in b) are offset for clarity.}
	\label{FigMT1}
\end{figure}



Figure~\ref{FigT1}(a) shows the sheet resistivity of the investigated samples, \#A, \#B, and \#C, at $T = \Kelvin{4.2}$ as a function of the back gate voltage $U_\mathrm{BG}$. For $U_\mathrm{TG} = U_\mathrm{TG1} = U_\mathrm{TG2} = \SI{0}{\volt}$ (black curve) a clear single maximum, corresponding to the charge neutrality point (CNP), is depicted in the response.
For sample \#C the sheet resistivity maximum of $\rho = \SI{0.54}{\kilo\ohm/\mathrm{sq}}$ is located at $U^\mathrm{max}_\mathrm{BG} = \SI{-0.15}{\volt}$ yielding a negligibly low doping of the structure. Samples \#A ($U^\mathrm{max}_\mathrm{BG} = \SI{-16}{\volt}$, $\rho = \SI{0.78}{\kilo\ohm/\mathrm{sq}}$) and \#B ($U^\mathrm{max}_\mathrm{BG} = \SI{-1}{\volt}$, $\rho = \SI{2.65}{\kilo\ohm/\mathrm{sq}}$) follow the same overall behavior, however, the former shows a noticably larger shift of the resistivity maximum.
This negatively shifted CNP in sample \#A reveals a high residual $n$-type doping. Since measurements were also performed at room temperature, the uncoated edges near the contacts may have attracted adsorbates changing the doping of the graphene structure. Hence, in the following, the applied back gate voltage will be presented as an effective voltage with $U_\mathrm{BG, eff} = U_\mathrm{BG} - U^{\mathrm{max}, m}_{\mathrm{BG}}$, where $m$ marks the sample indication.

The application of the same voltages $U_\mathrm{TG1} = U_\mathrm{TG2} \neq 0$ to the top gates, results in a substantial change of the resistivity response, see blue and red curves in Fig.~\ref{FigT1} I-III(a). Here, two peaks are apparent: the main Dirac peak near zero and a satellite peak at a positive (negative) back gate voltage in case of negative (positive) top gate bias. This additional resistivity maxima may be understood qualitatively in an extended capacitor model \cite{Castro2010,Oostinga2007}.

While the back gate acts on the entire graphene flake, the top gates only couple to the graphene regions underneath, shifting the CNP only in those regions. When the top gates are grounded, apart from a small contribution owing to the work function difference between the metallic top gate stripes and graphene, this leads to a single resistivity maximum, black curves in Fig.~\ref{FigT1} I-III(a). By contrast, when the top gate stripes are connected to a finite potential, the carrier concentration in the regions directly below is shifted, resulting in a double resistivity peak in the back gate sweep. This is also confirmed by Fig.~\ref{FigT1} I-III(b), where the resistivity $\rho$ is illustrated as a function of the applied top gate bias, while keeping both, back gate and one top gate, at zero and sweeping the other as indicated in the figures.

Figure~\ref{FigMT1}(a) shows the sheet resistivity of sample \#B in dependence on a perpendicular magnetic field for various effective back gate voltages ranging from 6 to \SI{11}{\volt} keeping the top gates at zero volts. Due to a two-point measurement geometry, the plotted resistivity is a superposition of longitudinal, Hall, and contact resistance contributions. To analyse and compare with the obtained photovoltage data at non-zero magnetic fields the Hall contribution was subtracted, providing the oscillatory part $\Delta\rho$ plotted in panel Fig.~\ref{FigMT1}(b). The resistivity decreases with increasing $U_\mathrm{BG, eff}$ and the Shubnikov-de Haas oscillations (SdHO) are shifted towards higher fields, see Fig.~\ref{FigMT1}. The carrier concentration is obtained by taking the fast Fourier transform of the SdHO, giving the gate-coupling factor $\alpha^\prime \approx \SI{6e10}{\centi\meter^{-2}\volt^{-1}}$ for the corresponding sample. 
We may conclude, that the discussed transport properties show that all samples are comparable bilayer structures of high quality demonstrating ranges similar to those found in several theoretical and experimental publications \cite{Monteverde2010,Woo2019,Avsar2016,Li2011}.

\section{Additional photovoltage measurements}
\label{Additional_measurements}

\subsection{Dependence of the photovoltage on the lateral asymmetry parameter}

\begin{figure}
	\centering
	\includegraphics[width=0.95\linewidth]{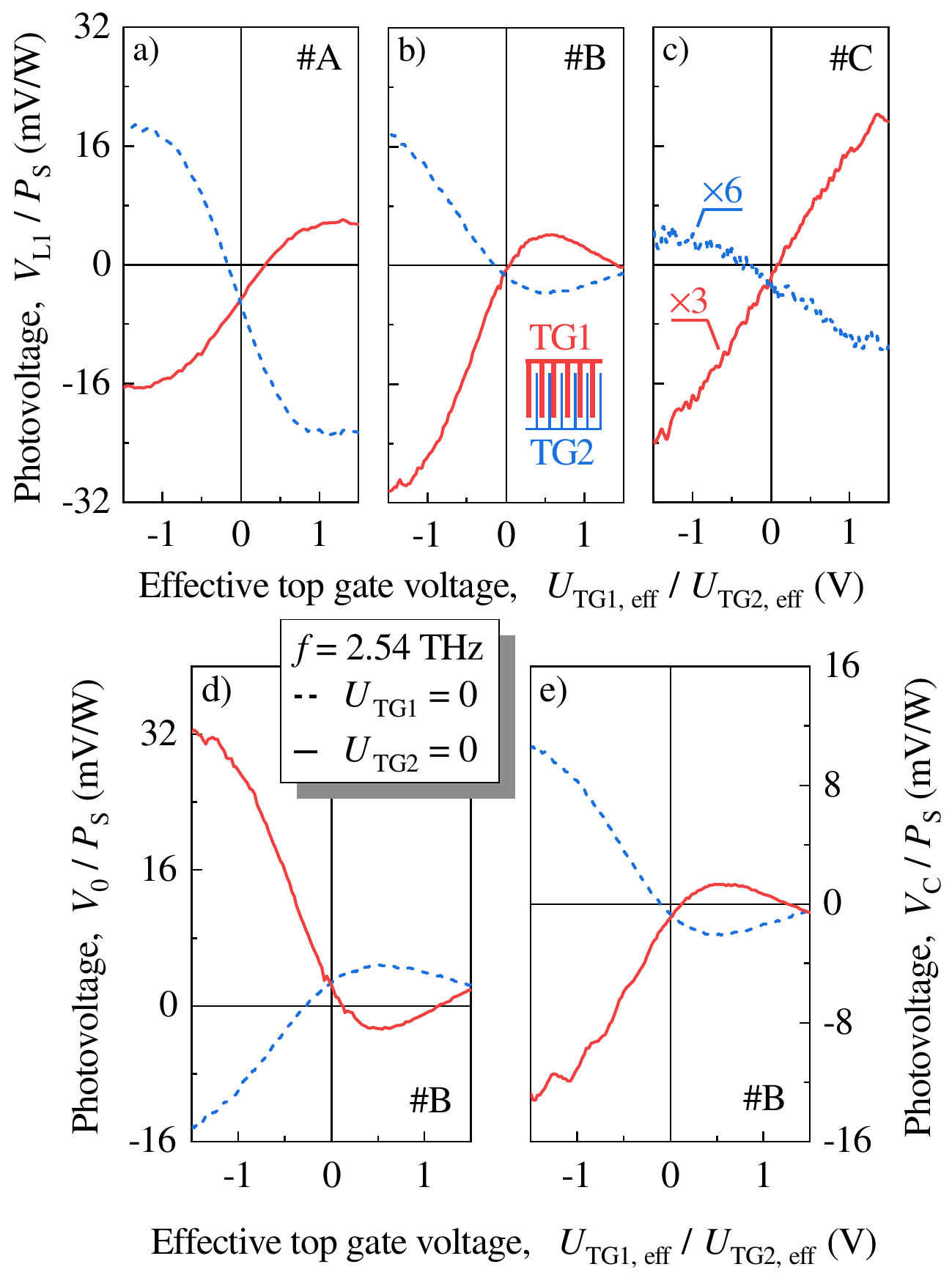}
	\caption{Amplitude of ratchet photovoltage normalized on radiation power $P_\mathrm{S}$ as a function of the effective top gate voltage $U_\mathrm{TG1}$ ($U_\mathrm{TG2}$) shown as red solid (blue dashed) curves measured for grounded top gate TG2 (TG1). Panels (a)-(c): The linear ratchet photovoltage  $V_\mathrm{L1}/P_\mathrm{S}$ measured in samples \#A ($T = 300$~K), \#B ($T = 150$~K), and \#C ($T = 150$~K), respectively. The data for sample \#A and \#C are presented for $U_{\rm BG, eff} = 0$ and for sample \#B for $U_{\rm BG, eff} = 1$~V. Panel (d): the polarization-independent ratchet signal $V_0/P_\mathrm{S}$ measured in sample \#B at $T=150$~K. Panel (e): Circular ratchet response obtained for sample \#B at $T=150$~K and $U_{\rm BG, eff} = 1$~V. The inset in panel (a) shows the DGG stucture. The color code corresponds to the applied top gates.
	}
	\label{FigR2}
\end{figure}

The ratchet behavior was observed for all samples investigated in this work. It is confirmed by the observation that $V_{\rm ph} \propto \Xi$. Figures~\ref{FigR2}(a), (b), and (c) show the results on the linear ratchet effect obtained for samples \#A, \#B and \#C, respectively. The polarization-independent and circular component are depicted in panels (d) and (e). To extract $V_0$, $V_\mathrm{L1},$ and $V_C$ from the total photoresponse we exploited different behaviors of individual contributions to the ratchet effect on radiation's polarization. The photovoltage was measured for $\alpha = 0$, $\Degree{90}$ and $\varphi = \Degree{45}(\sigma^+)$, $\Degree{135}(\sigma^-)$ to calculate the curves in Fig.~\ref{FigR2} accordingly [see Eq.~(\ref{linear}) and Eq.~(\ref{circular}) of the main text]. These figures reveal the presence of the built-in asymmetry and, clearly demonstrate that the inversion of $\Xi$, obtained either by the change of polarity of the gate voltage applied to one of the gates or by exchange of the biased gates, yield opposite signs of the photoresponse. In samples \#A and \#B featuring the same top gates stripe widths and spacings, we obtained very similar magnitudes of signals, whereas in sample \#C with two times narrower stripes and spacings ($d_1/d_2 = 0.5/\SI{0.25}{\micro\meter}$ and $a_1/a_2 = 1/\SI{0.25}{\micro\meter}$)	the amplitude of the ratchet voltage was  several times less, see Fig.~\ref{FigR2}(c).

\subsection{Backgate voltage dependence of the ratchet current}

\begin{figure}
	\centering
	\includegraphics[width=\linewidth]{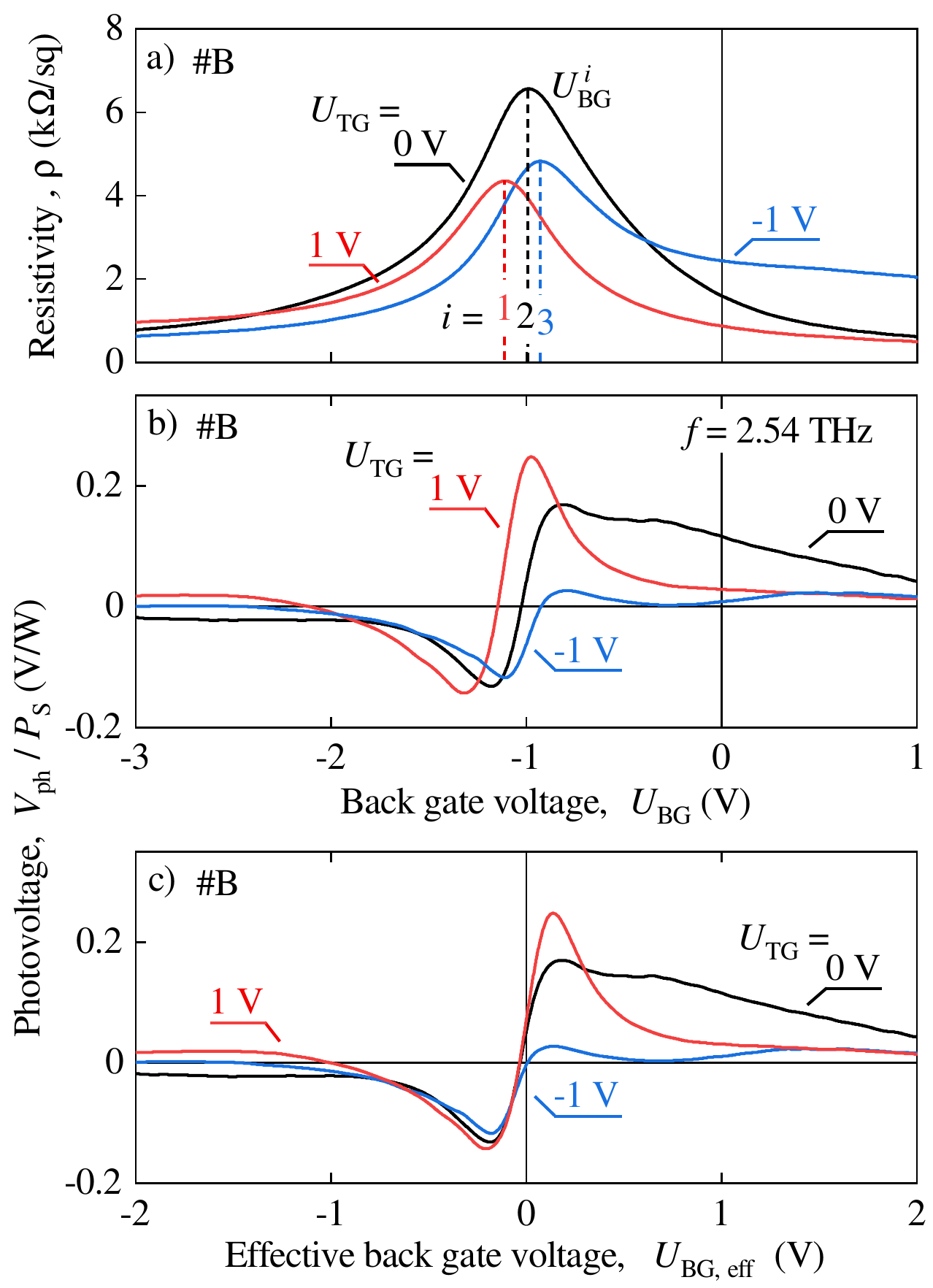}
	\caption{Back gate voltage dependencies of resistivity and photosignal measured in sample \#B at $T=4.2$~K and three values of top gate voltage $U_{\rm TG}= U_{\rm TG1}= U_{\rm TG2}$. Panel (a): Back gate voltage dependence of the two-terminal sheet resistivity. $U^i_\mathrm{BG}$ with $i$ = \{1,2,3\} labels back gate voltage of resistivity maxima. Panel (b): Back gate voltage dependence of the photosignal obtained for $\alpha = 90^\circ$. Panel (c): The curves replotted as a function of the effective back gated voltage calculated after $U_{\rm BG, eff} = U_{\rm BG} - U^i_\mathrm{BG}$.}
	\label{FigR3}
\end{figure}

At low temperatures the ratchet effect shows a sign-alternating behavior with enhanced magnitude in the vicinity of the CNP (see Fig.~\ref{FigR3}). A very similar behavior is also observed varying the top gate voltages (see  Fig.~\ref{FigR5}). Figs.~\ref{FigR3}(a) and (b) show sample's resistivity and the ratchet photosignal excited by linear polarized radiation as a function of $U_{\rm BG}$, respectively. For equally biased gates, $U_{\rm TG}= U_{\rm TG1}= U_{\rm TG2}$, we observed that the photovoltage inverses its sign close to the CNP and is strongly enhanced in its vicinity. To visualize it, in Fig.~\ref{FigR3}(c) the voltage $U^i_\mathrm{BG}$ corresponding to the maximum resistivity was subtracted from the back gate voltage $U_{\rm BG}$ and the data were plotted as a function of $U_{\rm BG, eff}$. Our analysis shows that this behavior can roughly be described by the first derivative of the conductance. Such behavior has been reported for the ratchet effect in monolayer graphene~\cite{Hubmann2020} and graphene based field-effect transistors~\cite{Vicarelli2012}. On the one hand, it is well known that the broad band rectification of terahertz radiation in an FET channel is proportional to the so-called FET factor given by $G^{-1} \frac{dG}{dV_G}$ and on the other, in graphene it can additionally be caused by enhanced plasmonic rectification of terahertz radiation in graphene structures towards the charge neutrality point suggested in Ref.~\cite{Fateev2019}. A detailed discussion of these effects is out of scope for this paper.


\begin{widetext}
\section{Derivation of analytical expression  showing crossover from the
hydrodynamic to the drift-diffusion regime}
\label{App_theory}

Here, we derive analytical equation for dc current valid for arbitrary relation
between $\gamma$ and $\gamma_{\rm ee}$ and between  $s$ and $v_F.$
Using dimensionless variables $$ n=\delta N/N_{0},~N/N_{0}=1+n,~j=J/N_{0},~\pi=\Pi/N_{0},$$ we rewrite Eqs.~\eqref{system0} as follows
\begin{align}
& \frac{\partial n}{\partial t}+\nabla_{\rm x} j=0, \\
& \frac{\partial j}{\partial t}+j\gamma+\frac{1}{2}\nabla_{\rm x}\left\{\frac{v_{\rm F}^2}{2}(1+n)^2 + \frac{j^2}{1+n}\right\} + \nabla_{\rm x}\pi=f  (1+n), \\
& \frac{\partial \pi}{\partial t}+(\gamma+\gamma_{\rm ee})\pi=f j+\gamma_{\rm ee} \frac{j^2}{2(1+n)}.
\end{align}

Now, we  take into account the plasmonic collective force and replace
\be  f \to f-s^2 \partial_{x} n, \ee
where $f$ in the r.h.s of this equation  corresponds to external force.
Then, we arrive at the following set of equations:
\begin{align}
& \frac{\partial n}{\partial t}+\nabla_{\rm x} j=0,
\label{n-App}
\\
& \frac{\partial j}{\partial t}+j\gamma+\left(\frac{v_{\rm F}^{2}}{2}+s^2\right)\nabla_{\rm x} n+\nabla_{\rm x}\pi=A,
\label{j-App}
\\
& \frac{\partial \pi}{\partial t}+(\gamma+\gamma_{\rm ee})\pi=B.
\label{pi-App}
\end{align}
The  terms  $A$ and  $B$ entering r.h.s of these equations read
\begin{align}
& A=f+f\cdot n -s^2 n\partial_{x} n-\frac{v_{\rm F}^2}{4}\partial_{\rm x}\left[n^2+\frac{2 j^2}{v_{\rm F}^{2}(1+n)} \right],
\label{A}
\\
& B=\left(f-s^2 \partial_{x} n\right) j+\gamma_{\rm ee}  \frac{j^2}{2(1+n)}.
\label{B}
\end{align}
There is a linear in $f$ term here and also a number of nonlinear terms. Following  method used in Ref.~\cite{Rozhansky2015} we will solve this system perturbatively with respect to
\be
f= -\frac{e E_0}{m} \left[ 1+ h \cos(q x+ \varphi)\right] \cos\omega t + \frac{1}{m} \frac{\textrm{d}U_0}{\textrm{d}x}.
\ee
We will use the  notation $a^{(i,j)} \propto (E_0)^i (U_0)^j $ for any quantity $a$ calculated  in the $i-$th order in $E_0$ and $j-$th order in $U_0.$ For example,   $n^{(1,1)}$  denotes concentration, calculated in  the first order  with respect to $U_0$ and  first order with respect
$E_0.$  The nonzero  response to the current appears in the order $(2,1)$ and is proportional  to asymmetry factor $\Xi.$  Averaging Eq.~\eqref{j-App}
over  $t$ and $x$ and taking into account Eq.~\eqref{A},
we find that  dc current  is given by
\be
j_{\rm dc}={\gamma^{-1}}  \langle f n \rangle_{t,x}.
\label{jdc}
\ee
Hence, we need to calculate  $f n$ in the order $(2,1).$

Importantly,  at  each iteration step,  one can use values of
parameters  $A$ and $B$ entering r.h.s. of Eqs.~\eqref{n-App}, \eqref{j-App}, and \eqref{pi-App}    found in the previous step.    Let us clarify  this point in more detail.  At the first step, we  simply neglect all nonlinear terms in $A$ and $B,$  i.e. writing  $A=f,~B=0.$ The force $f$ can be presented as $$f= f_{10}+\tilde f_{10} +f_{01},$$ where
$$
f_{10}=- \frac{eE_{0}}{2m}e^{-i\omega t} +c.c.,
\quad \tilde f_{10}=- \frac{eE_{0}}{2m}h\cos(q x + \phi) e^{-i\omega t} +c.c.,
\quad f_{01}=\frac{1}{m} \frac{\textrm{d}U_0}{\textrm{d}x}.$$
Here and in what follows, we use notation $ \tilde f_{10} $ and $  f_{10} $
  to  distinguish between $x-$dependent and  $x-$independent  terms of the same order.
Solving then  Eqs.~\eqref{n-App}, \eqref{j-App}, and \eqref{pi-App}, we find $n,j$ and $\pi$ in the orders $(1,0)$ and $(0,1).$ Substituting these solutions into  quadratic nonlinear terms in $A$ an $B,$  we find sources for orders $(1,1), (2,0),$  and  $ (0,2). $     The terms of the  needed order  $(2,1)$ appear on the next iteration step. On each iteration step, we solve  Eqs.~\eqref{n-App}, \eqref{j-App}, and \eqref{pi-App}           with given  coefficients  $A$ and $B.$ It is convenient to  write this solution in the formal operator form
 \be
\label{iteraton-step}
\begin{aligned}
& n_{ij}=\frac{ 1 }{(\gamma+\gamma_{\rm ee}+\p_t )(\gamma \p _t + \p_t^2 -s_0^2 \p_x^2)} \left[ -(\gamma+\gamma_{\rm ee} + \p_t)\p_x A_{ij}  + \p_x^2 B_{ij}  \right],
\\
& j_{ij}=
\quad\frac{ 1 }{(\gamma+\gamma_{\rm ee}+\p_t )(\gamma \p _t + \p_t^2 -s_0^2 \p_x^2)} \left[ (\gamma+\gamma_{\rm ee} + \p_t)\p_t A_{ij}  - \p_x \p_t B_{ij}  \right],
\\
&\pi_{ij}=\frac{1}{\gamma+\gamma_{\rm ee} +\p_t} B_{ij}.
\end{aligned}
\ee
Here  $A_{ij}= A_{ij} (x,t) $ and $B_{ij}=B_{ij} (x,t) $ are the r.h.s.
of Eqs.~\eqref{j-App} and \eqref{pi-App}, respectively, in the order $(i,j).$ Due to nonlinearity of the problem,  characteristic frequencies and
wave vectors entering  Fourier transform of  $A_{ij},B_{ij}$ at the each step of iteration  are  given by harmonics of the $\omega$ and $q,$ respectively: $M\omega, K q$,where $M$ and $K$ are integer numbers.   For $M=\pm 1$ and $K=\pm, $  Eq.~\eqref{iteraton-step} shows plasmonic resonance at $\omega\approx\omega_{q}.$

Using Eq.~\eqref{jdc}, we find that in the lowest nonzero perturbation order, $(2,1),$  dc current is given by the sum of two contributions:
\be
J_{\rm dc,x}\approx\ e N_{0} \gamma^{-1}\langle f~ n\rangle_{\rm t,x}^{2,1}= J_{\rm dc,x}^{\rm I}+J_{\rm dc,x}^{\rm II},
\label{Jdc-(tot)}
\ee
where
\be
J_{\rm dc,x}^{\rm I} \approx e N_{0} \gamma^{-1}\langle(f_{10}+\tilde{f}_{10})n_{11}\rangle_{\rm t,x},
\label{Jdc-(21)}
\ee
and
\be
J_{\rm dc,x}^{\rm II}\approx e N_{0} \gamma^{-1}\langle f_{01}n_{20}\rangle_{\rm t,x}.
\label{Jdc-(20)}
\ee
In order to find  $n_{11}$ and $n_{20},$ we need to do two iterations.
At the first iteration step, we  find:
\begin{align}
& n_{10}=0, \quad j_{10}=-\frac{e E_{0}}{2m}\frac{i}{\omega+i\gamma}e^{-i\omega t}+c.c.,\\
& n_{01}=\frac{e U_{0}}{m s_{0}^2}\cos{(q x)}, \quad j_{01}=0,\\
& \tilde{n}_{10}=\frac{e E_0 h}{2 m}\frac{q \sin{(q x)} e^{-i \omega t}}{\omega(\omega+i \gamma)-\omega_{q}^2}+c.c., \quad \tilde{j}_{10}=-\frac{e E_0 h}{2 m}\frac{i \omega \cos{(q x+ \phi)}e^{-i \omega t}}{\omega(\omega+i \gamma)-\omega_{q}^2}.
\end{align}
Using these equations, one can find sources
\be
\begin{aligned}
& A_{11}=(f_{10}+\tilde{f}_{10})n_{01}+f_{01}\tilde{n}_{10}-s_{0}^2 \partial_{x}\{\tilde{n}_{10}n_{01}\}, \\
& B_{11}=(\tilde{j}_{10}+j_{10})(f_{01}-s^2 \partial_{x} n_{01}), \\
& A_{ 20}=f_{10}\tilde{n}_{10}-\partial_{x}\{j_{10}\tilde{j}_{10}\},\\
& B_{20}=\gamma_{\rm ee}j_{10}\tilde{j}_{10}+f_{10}\tilde{j}_{10}+\tilde{f}_{10}j_{10}-s^2j_{10}\partial_{x}\tilde{n}_{10}.
\end{aligned}
\label{AB-20-11}
\ee
which  allow to   make the next iteration step by solving Eqs.~\eqref{iteraton-step} with $(A_{11}, B_{11}),$  and   $(A_{20}, B_{20})$ in the r.h.s..  Doing so, we  can find $n_{11}$ and $n_{20},$ respectively.  In Eqs.~\eqref{AB-20-11},  we neglected terms, prohibited by symmetry consideration,  for example, $\tilde f_{10} \tilde n_{10}.$   Equations  Eqs.~\eqref{AB-20-11} contains also terms  with product of two spatially oscillating function.  For example, $ \tilde f_{10} n_{01} \propto \cos(q x + \varphi) \cos(q x) $ (and other similar terms), which contain
0th and $2 q$ spatial harmonics. Zero harmonics gives zero contribution to $n_{11},n_{20}$ as follows from  Eq.~\eqref{iteraton-step}.
        Second harmonics yields  $n_{11},~n_{20} \propto \exp (2 i q x ).$ These terms drop out after spatial averaging in Eqs.~ \eqref{Jdc-(21)} and  \eqref{Jdc-(20)}.
Leaving terms, which give non-zero contributions, after simple algebra we find
\begin{align}
& A_{11}=f_{10}n_{01}=-\frac{e E_{0} U_0}{2 m^2 s_{0}^2}\cos{(q x)}e^{-i\omega t}+c.c.,\\
& B_{11}=
j_{10}(f_{01}-s^2 \partial_{x} n_{01})
=\frac{q e E_{0} U_{0}}{2 m^2}\left(1-\frac{s^2}{s_{0}^2}\right)\sin{(q x)}\frac{i e^{-i\omega t}}{\omega+i\gamma}+c.c.,\\
& A_{20}=\langle f_{10}\tilde{n}_{10}-\partial_{x}\{j_{10}\tilde{j}_{10}\}\rangle_{t}=\frac{q h e^2 E_{0}^2}{4 m^2}\sin{(q x+\phi)}\left\{\frac{i \gamma}{(\omega-i\gamma)(\omega^2+i\gamma\omega-\omega_{q}^2)}+c.c.\right\},\\
& B_{20}=\langle \gamma_{\rm ee}j_{10}\tilde{j}_{10}+f_{10}\tilde{j}_{10}+\tilde{f}_{10}j_{10}-s^2j_{10}\partial_{x}\tilde{n}_{10}\rangle_{t}=\frac{h e^2 E_{0}^2}{4 m^2}\cos{(q x + \phi)}\left\{\frac{\omega(\gamma_{\rm ee}+2\gamma)+iq^2(s_{0}^2-s^2)}{(\omega-i\gamma)(\omega^2+i\gamma\omega-\omega_{q}^2)}+c.c.\right\}.
\end{align}
Having in mind that Eq.~\eqref{Jdc-(20)} contains averaging over time and that $f_{01}$ does not depend on time,we averaged $A_{20}$ and $B_{2,0}.$
Substituting these formulas into Eq.~\eqref{iteraton-step} we find:
\begin{align}
 n_{11}&=\frac{\partial_{x} A_{11} (i\omega-\gamma-\gamma_{ ee})+\partial_{x}^2 B_{ 11}}{(\gamma+\gamma_{ ee}-i\omega)(\omega_{q}^2-\omega(\omega+i \gamma))}+c.c.,
 \\ &=\frac{e q E_{0} V_{0}}{2 m^2 s_{0}^2}\sin{(q x)}\left\{\frac{i\omega-(\gamma+\gamma_{\rm ee})-q^2(s_{0}^2-s^2)\frac{i}{\omega+i\gamma}}{(\gamma+\gamma_{\rm ee}-i\omega)(\omega_{q}^2-\omega(\omega+i\gamma))}e^{-i\omega t}+c.c.\right\},
 \end{align}

\begin{align}
 n_{20} &=\frac{-\partial_{x} A_{20}(\gamma+\gamma_{\rm ee})+\partial_{x}^2 B_{20}}{\omega_{q}^2(\gamma+\gamma_{\rm ee})}
\\
&=\frac{e^2 E_{0}^2 h}{4 m^2}\cos{(q x+\phi)}\left\{\frac{-i\gamma(\gamma+\gamma_{\rm ee})-\omega(\gamma_{\rm ee}+2\gamma)+i q^2(s_{0}^2-s^2)}{s_{0}^2 (\gamma+\gamma_{\rm ee})(\omega-i\gamma)(\omega^2+i \gamma \omega-\omega_{q}^2)}+c.c.\right\}.
\end{align}
Next, we do the final  step, namely, substitute $n_{11}$ and $n_{20}$ into Eqs.~\eqref{Jdc-(21)} and \eqref{Jdc-(20)}  and find two terms contributing  to dc current:
\begin{align}
& J_{\rm dc,x}^{\rm I}=  \frac{C_1s_0^2}{2 \gamma s^2}~\frac{i\omega-(\gamma+\gamma_{\rm ee})+q^2(s^2-s_{0}^2)\frac{i}{\omega+i\gamma}}{(\gamma+\gamma_{\rm ee}-i\omega)(\omega_{q}^2-\omega(\omega+i\gamma))}+c.c.,
\label{j-final-n11}
\\
\label{j-final-n20}
& J_{\rm dc,x}^{\rm II}=\frac{C_1s_0^2}{2 \gamma s^2}~\frac{- i \gamma(\gamma+\gamma_{\rm ee})-[\omega(\gamma_{\rm ee}+2\gamma)+iq^2(s_{0}^2-s^2)]}{(\gamma+\gamma_{\rm ee})(\omega-i\gamma)(\omega(\omega+i\gamma)-\omega_{q}^2)}+c.c.,
\end{align}
where $C_1$ is given by Eq.~\eqref{C0}. Summing up these equations and skipping  terms
$\propto (s_0^2 -s^2),$ we reproduce Eq.~\eqref{final-current-large-s} of the main text
\end{widetext}

\newpage
\bibliography{references}

\end{document}